\documentclass[a4paper,11pt]{article}
\usepackage{jcappub} 

\usepackage{amsmath}	
\usepackage{bm}

\usepackage{multirow}
\usepackage{booktabs}
\usepackage{mathtools}
\usepackage{array}
\usepackage{arydshln}
\usepackage{changepage} 

\usepackage{graphicx}	
\usepackage{subcaption}

\usepackage{aasmacros}

\usepackage{ulem}

\makeatletter
\newcommand\notsotiny{\@setfontsize\notsotiny\@vipt\@viipt}
\makeatother

\newcommand{\mymatrix}[1]{\mathbf{#1}}
\newcommand{\myset}[1]{{#1}}
\newcommand{\polarization}{}
\newcommand{\myvector}[1]{\mathbf{#1}}

\title{\boldmath Addressing Synchrotron Challenges for CMB Observations: ELFS-SA Collaboration for Robust Foreground Removal}

\author[1,2]{E. de la Hoz,}
\author[3,4]{A. Mennella,}
\author[5]{K. Arnold,}
\author[6]{C. Baccigalupi,}
\author[7]{A. J. Banday,}
\author[8]{R. B. Barreiro,}
\author[9]{D. Barron,}
\author[3,4]{M. Bersanelli,}
\author[8]{F. J. Casas,}
\author[5]{S. Casey,}
\author[3,4]{C. Franceschet,}
\author[10]{M. E. Jones,}
\author[11]{R. T. Genóva-Santos,}
\author[11]{R. Hoyland,}
\author[2]{A. T. Lee,}
\author[8]{E. Martinez-Gonzalez,}
\author[3]{F. Montonati,}
\author[11]{J.-A. Rubiño-Martín,}
\author[10]{A. C. Taylor,}
\author[8]{and P. Vielva}

\affiliation[1]{CNRS-UCB International Research Laboratory, Centre Pierre Binétruy, IRL2007, CPB-IN2P3, Berkeley, CA 94720, USA}
\affiliation[2]{University of Berkeley, Department of Physics, 366 Physics North MC 7300 Berkeley, CA, 94720-7300, USA}
\affiliation[3]{University of Milan, Department of Physics, Via Celoria 16, Milan, Italy}
\affiliation[4]{INFN Milan section, Via Celoria 16, Milan, Italy}
\affiliation[5]{University of California at San Diego, 5998 Alcala Park, San Diego, CA 92110-8001, USA}
\affiliation[6]{SISSA, Via Bonomea 265, 34136 Trieste, Italy}
\affiliation[7]{IRAP, Université de Toulouse, CNRS, CNES, UPS, Toulouse, France}
\affiliation[8]{IFCA (CSIC-UC), Avenida de los Castros s/n, 39005 Santander, Cantabria, Spain}
\affiliation[9]{The University of New Mexico, Dept. of Physics and Astronomy, Albuquerque, NM 87131, USA}
\affiliation[10]{University of Oxford, Department of Physics, Denys Wilkinson Building, Keble Road, Oxford, OX1 3RH, UK}
\affiliation[11]{IAC, Calle Vía Láctea s/n, E-38205 La Laguna, Tenerife, Spain}

\emailAdd{delahoz@berkeley.edu}

\abstract{
Upcoming cosmic microwave background (CMB) experiments aim to detect primordial gravitational waves with unprecedented sensitivity. Effective foreground removal is essential to avoid biases in the measurement of the tensor-to-scalar ratio ($r$) in this high-precision regime. Recent analyses highlight the unexpected complexity of synchrotron emission at low frequencies, underscoring the need for more sensitive low-frequency data. To address this challenge, the European Low-Frequency Survey (ELFS) initiative and the Simons Array collaboration propose installing two European low-frequency receivers on one of the Simons Array telescopes. 
These receivers will enable measurements in the Southern Hemisphere between $6$ and $20$\,GHz, complementary to those of current and proposed experiments targeting the measurement of cosmological gravitational waves. 
In this work, we study the benefits of combining these low-frequency observations with a representative future CMB experiment operating from the Southern Hemisphere. 
We find that the extra information can improve the knowledge of the underlying synchrotron spectral energy distribution (SED), with positive impacts on the robustness of measurement of the tensor-to-scalar ratio, $r$, against the complexity of low-frequency foregrounds.}

\begin{document}
\maketitle
\flushbottom


\section{Introduction} \label{sec:outline}

Upcoming CMB experiments aiming to detect primordial gravitational waves (PGWs) \cite{guth1981inflationary,linde1982new}, are set to achieve unprecedented sensitivity levels. In this scenario of high-precision measurements, one of the most important concerns is the effective mitigation of foreground contamination. This is essential to ensure an unbiased measurement of the tensor-to-scalar ratio ($r$) \cite{fg_tensor_to_scalar_ratio}, the fundamental cosmological parameter that characterizes the relative power of the primordial gravitational waves with respect to their scalar counterpart.

Within the microwave frequency spectrum, the dominant polarized components of the foreground consist of thermal dust and synchrotron emissions, observed at higher and lower frequencies, respectively. Significant attention has already been directed towards the study of Galactic thermal dust, particularly after the false detection of CMB $B$-mode polarization by BICEP, which was subsequently attributed to dust emission \cite{bicep_nondetection_r,bicep_planck_dust_r}. A thorough understanding of the properties of dust is essential to prevent false detection claims in the future, where dust could be mistaken for primordial signals if an incorrect dust model is assumed \cite{dust_mismodelling_remazeilles,dust_mismodelling_hensley}.

Thus, upcoming CMB experiments such as Simons Observatory (SO) \cite{SO}, \textit{LiteBIRD} \cite{LiteBIRD_ptep}, and Stage-4  CMB experiments like \cite{CMB-S4}, whose main goal is to detect PGWs, have been carefully designed to provide extensive frequency coverage, with an emphasis on the higher frequency bands. However, there has been less emphasis on synchrotron radiation influencing the design of current CMB experiments, which generally exclude the low-frequency region where synchrotron is the most dominant component ($\nu \lesssim 20$\,GHz). This is partly because synchrotron emission has generally been assumed to follow a simple spectral behaviour, i.e. a simple power-law, which requires fewer observational constraints, and also because of the technical difficulties of observing with bolometer arrays at lower frequencies.

However, recent  analyses that incorporate low-frequency data ($\sim 2$-$20$\,GHz) from ground-based experiments \cite{MFI,SPASS,C-BASSPol} have revealed that the dominant synchrotron emission at low frequencies is more complex than expected. Experimental evidence reveals larger spatial variations of the synchrotron spectral index ($\beta_s$), along with indications of possible deviations from the simplistic power--law model, such as the presence of curvature \cite{planck_int_low_diff_fg,synch_galprop_model}. These empirical results find theoretical support in various models. For example, cosmic-ray transport models predict curvature of the synchrotron spectrum in the microwave frequency regime \cite{planck_int_low_diff_fg,synch_galprop_model}. Furthermore, it is expected that $\beta_s$ will vary with position and frequency due to factors such as synchrotron ageing and self-absorption \cite{synchrotron_aging,synchrotron_self_absorption}. Also, the presence of multiple distinct groups of relativistic electrons along a single observational line of sight, each characterized by a distinct energy spectrum, results in a composite observed spectrum that need not follow a power law.

In order to understand the spectral energy distribution (SED) of the synchrotron and to avoid potential biases in $r$ arising from incorrect or insufficient removal of this emission, it is crucial to acquire more sensitive low-frequency data. In this regard, we present a collaborative effort between the European Low-Frequency Survey (ELFS)  initiative \cite{ELFS} and the Simons Array (SA) collaboration \cite{SA}. We propose a new experiment ELFS on SA (ELFS/SA) to measure this frequency regime using available resources. Our plan involves installing two available European low-frequency receivers on one of the Simons Array telescopes, enabling measurements of the Southern Hemisphere sky between $6$ and $20$\,GHz. This experiment will allow us to determine whether the synchrotron SED can be adequately modelled as a power law, or if it is more complex. If it is more complex, the new data will also allow us to better clean the synchrotron foregrounds from the CMB.

In this paper, we assess whether this experiment can achieve these goals by performing a comprehensive forecasting analysis which combines the low-frequency sensitivity of ELFS/SA with the expected performance of ground-based CMB experiments in the coming decade. First, we evaluate what can be learned about the synchrotron SED with and without the inclusion of low-frequency data, and second, if the synchrotron SED happens to be more complex than a power-law model, we study the impact of this extra information on simulated CMB measurements. We use available estimates of the sensitivity of SO as a template for assessing the likely sensitivity of upcoming high-frequency CMB observations, but this analysis is relevant to all future experiments that will achieve high sensitivity at the main CMB frequencies, including \textit{LiteBIRD} and Stage-4 CMB experiments.

This paper is structured as follows. In Section~\ref{sec:synchrotron} we review the status of synchrotron spectral modelling. Section~\ref{sec:elfs_on_sa} presents the ELFS on SA proposal. In Section~\ref{sec:simulations} we describe the simulations used in the analysis while Section~\ref{sec:methodology} outlines the component separation method used to analyze the data, and the cosmological likelihood to estimate the tensor-to-scalar ratio. In Section~\ref{sec:results} we present the results of combining the ELFS-SA instrument with high-frequency CMB data. Finally, Section~\ref{sec:conclusions} highlights the main conclusions from this study.  
\section{Synchrotron emission as a CMB foreground}
\label{sec:synchrotron}

Galactic synchrotron emission arises from cosmic rays (CR) spiraling around the Galactic magnetic field (GMF), serving as a crucial probe for understanding the interstellar medium (ISM). This emission is very important in astronomy not only because it places constraints on the GMF and the distribution of CRs but also because it presents challenges in cosmological analyses due to its overlap with key cosmological signals such as the CMB or the HI $21$-cm line.

A robust theoretical model of synchrotron emission would facilitate the accurate removal of this signal in cosmological analyses, thereby minimizing biases and reducing uncertainties in the recovered data. However, current models are only qualitatively compatible with observations, showing that the synchrotron spectrum steepens from $\beta_s \sim -2.5$ at $22$\,MHz to $\beta_s \sim -3.0$ above $23$\,GHz, \cite{model_radio_10MHz_100GHz,model_radio_45_408MHz}. This is a result of the degeneracy between the CR electron spectrum and the GMF. Breaking these degeneracies requires additional observations across a wide range of frequencies, as highlighted in \cite{Orlando2018}.

\begin{figure}
    \centering
    \includegraphics[width=.75\linewidth]{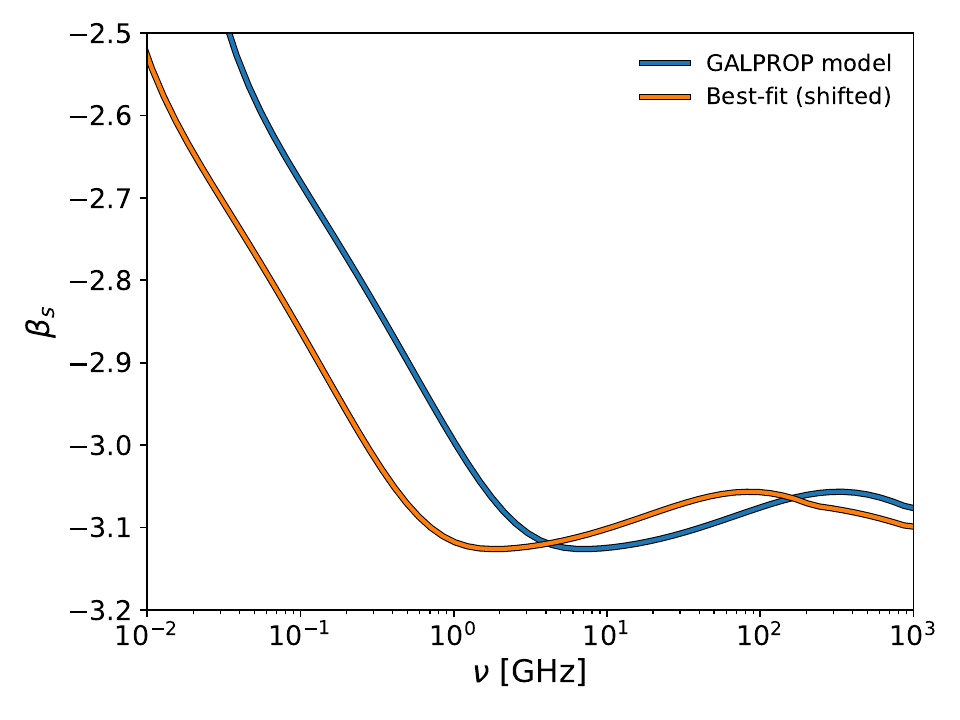}
    \caption{ Synchrotron SED used in the intensity component separation analysis, computed from GALPROP \cite{synch_galprop_model} (blue), and the best-fit SED obtained after \textit{Planck}'s Commander component separation allowing for a frequency shift (orange) \cite{planck_int_diff_fg}. 
    }
    \label{fig:beta_s_planck_galprop}
\end{figure}

Due to the complexity of theoretical modeling, the synchrotron SED is often modeled by a simple analytical function, whose parameters are derived from fitting observational data. Since the synchrotron depends on the magnetic field strength and CR energy, assuming that the CR distribution is well-described as a power law $N(E) \propto E^{s}$, the synchrotron spectrum is also modeled as a power-law:
\begin{equation}
    S\left(\mathbf{n}\right) = a_s\left(\mathbf{n}\right) \left(\dfrac{\nu}{\nu_s}\right)^{\beta_s\left(\mathbf{n}\right)}	\, ,
    \label{eq:powerlaw}
\end{equation}
where $a_s$ is the synchrotron intensity measured at frequency $\nu_s$ in the $\mathbf{n}$ direction, and $\beta_s$ is the spectral index. The spectral index $\beta_s$ is expected to exhibit both spatial and spectral variability, influenced by factors such as synchrotron ageing, self-absorption, and multiple populations along the line of sight. Given these complexities, a model incorporating a curvature parameter $c_s$ may offer a better fit: 
\begin{equation}
    S\left(\mathbf{n}\right) = a_s\left(\mathbf{n}\right) \left(\dfrac{\nu}{\nu_s}\right)^{\beta_s\left(\mathbf{n}\right) + c_s\left(\mathbf{n}\right)\log\left(\nu/23\,\mathrm{GHz}\right)}	\, ,
    \label{eq:powerlaw_curv}
\end{equation}
However, extracting synchrotron parameters from intensity multifrequency data remains challenging due to significant degeneracies with other low-frequency foregrounds, such as free-free emission and anomalous microwave emission. For example, the \textit{Planck} experiment was unable to fully resolve these degeneracies and relied on the theoretical synchrotron SED model from \cite{Orlando2013}, as shown in Fig.~\ref{fig:beta_s_planck_galprop}, to mitigate confusion with other low-frequency foregrounds \cite{PlanckCompSep2015,PlanckLowFreqFg2015}.

Polarization measurements yield a cleaner characterization of the synchrotron spectral index since it constitutes the dominant emission at frequencies below $\leq 70$ GHz. However, the current polarization maps from missions like WMAP and \textit{Planck} have low signal-to-noise ratios (S/N), which limit their accuracy in deriving spectral indices \cite{PlanckCompSep2018}. Foreground simulators like \texttt{PySM}\footnote{\url{https://github.com/galsci/pysm}.} \cite{pysm3} and \texttt{PSM}\footnote{\url{https://apc.u-paris.fr/~delabrou/PSM/psm.html}.} \cite{PSM} still rely on the pre-\textit{Planck} $\beta_s$ template from \cite{beta_s_template}. This model estimated the polarization fraction of the WMAP polarized K-band \cite{WMAP_9years}, assuming a bi-symmetrical spiral model of the GMF with a turbulent component following a $-5/3$ power law spectrum. From that result, $\beta_s$ was calculated by extrapolating the intensity at $23$\,GHz to Haslam’s $408$\,MHz map \cite{haslam}. While this template was useful at the time to perform forecast analysis with \textit{Planck}, it no longer provides a realistic representation of the sky.

Moreover, as noted earlier, the synchrotron SED steepens at higher frequencies, which can be better modeled using a curvature parameter, as described by Eq.~\ref{eq:powerlaw_curv}. In \cite{Curvature_arcade}, a steepening of the spectral index by $\Delta\beta=0.07$ per octave was observed, consistent with theoretical models of CR propagation in the GMF. A model that takes into account this steepening is included in \texttt{PySM} by extrapolating the curvature found in \cite{Curvature_arcade} from the ARCADE patch to the whole sky. Despite the added complexity to the SED, Eq.~\ref{eq:powerlaw_curv} does not capture the slight flattening that is predicted by CR rays models in the microwave regime, which could potentially make the search for PGW more complicated.

As stated above, our understanding of synchrotron radiation has since improved significantly with new data from ground-based experiments operating in the $2$–$20$ GHz range. They have revealed larger spatial variations in the synchrotron spectral index than previously expected as well as showing a preference for a steeper spectrum at $23$\,GHz compared to $2.3$\,GHz \cite{synchrotron_spass,synch_mfi}. As a result, \texttt{PySM} introduced an updated template for $\beta_s$, which accounts for this variability by applying a multiplicative factor of $1.6$ to the original template from \cite{beta_s_template}, with a mean shift from -$3$ to -$3.1$ and additional Gaussian small-scale fluctuations. Although this revised model is more statistically compatible with current observations, it still does not capture the full complexity of the synchrotron SED.

To explore different potential synchrotron scenarios, in this work we generate four sets of synchrotron simulations. The first and second sets are based on the \texttt{s5} and \texttt{s7} models respectively from \texttt{PySM}, which assume a simple power law and a power law with negative curvature with a pivot frequency of $23$\,GHz. For the third and fourth sets, we generate two different curvature map templates compatible with current statistics obtained from the analysis of the Southern Hemisphere sky. A comprehensive description of the latter simulations is given in Section~\ref{sec:simulations}.

\section{The ELFS on SA project}
\label{sec:elfs_on_sa}
The ELFS initiative is a long-term plan to deploy dedicated instruments and telescopes to produce a full-sky survey in the $5$--$100$\,GHz range with an angular resolution of $\sim 20$\,arcmin at $10$\,GHz, and sensitivity that will allow for $B$-mode extraction from data produced by current and future CMB experiments \cite{mennella-2024a}. ELFS/SA is the first stage of this project, being carried out in synergy with the SA collaboration. ELFS/SA will have two phases, associated with two instruments: a $5.5$--$11$\,GHz (Xband) receiver as phase $1$, and the $10$--$20$\,GHz QUIJOTE-MFI2 \cite{MFI2} as phase $2$. Both receivers will be installed sequentially in the Gregorian focus of one of the SA telescopes. Here, we describe the characteristics of the first instrument, which is being specially developed for this project. 

\subsection{Telescope}

The ELFS receivers will be fitted to one of the SA telescopes \cite{SA}. These are off-axis Gregorian telescopes which satisfy the Mizugushi-Dragone condition which results in a symmetrical, low-cross polarization beam equivalent to that from an on-axis design, but with no blockage effects. The central section of the primary mirror is a monolithic structure with a projected diameter of $2.5$\,m. This is surrounded by a set of panels that extend the paraboloidal shape to $3.5$\,m diameter. At the original millimeter wavelength operation of the telescope these panels constitute a guard band, but for centimetre wave operation they are sufficiently accurate to regard as part of the primary, and the feed horn described below is designed to illuminate the full $3.5$\,m aperture. The primary and secondary mirrors are surrounded by absorbing baffles which further reduce far-out sidelobes, and prevent any direct ground radiation from hitting either the secondary or the feed.

\subsection{The ELFS/SA Xband receiver} 
\begin{figure*}
\begin{center}
    \begin{tabular}{c} 
    \includegraphics[width=14cm]{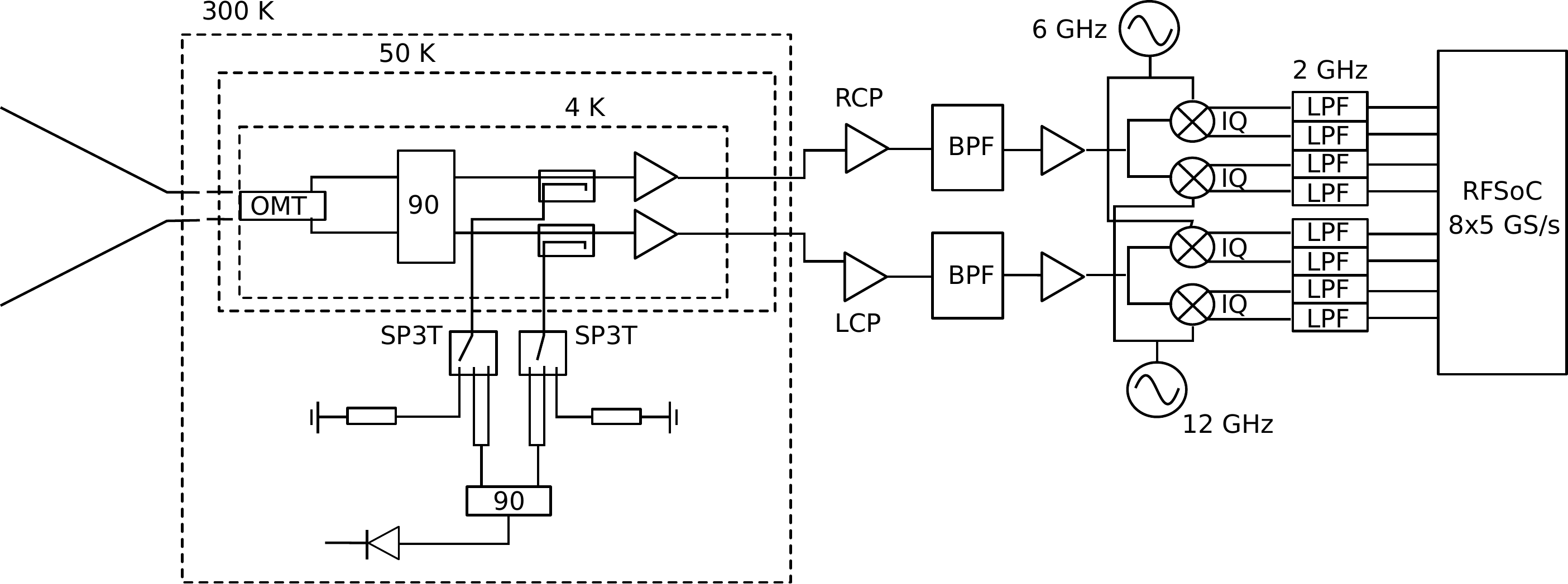}
    \end{tabular}
\end{center}
\caption{\label{fig_receiver_schematics}
        Schematic diagram of the ELFS/SA Xband receiver.}
\end{figure*}

Fig.~\ref{fig_receiver_schematics} shows a schematic view of the Xband receiver. The front end will be cooled to $4$\,K in a cryostat adapted from the C-BASS North receiver \cite{C-BASS, 2014MNRAS.438.2426K}, modified to accommodate a new $2$:$1$ bandwidth feed horn and orthomode transducer (OMT) in place of the $30$\,\% bandwidth horn and OMT used in C-BASS.

\begin{figure*}
\begin{center}
    \includegraphics[width=14cm]{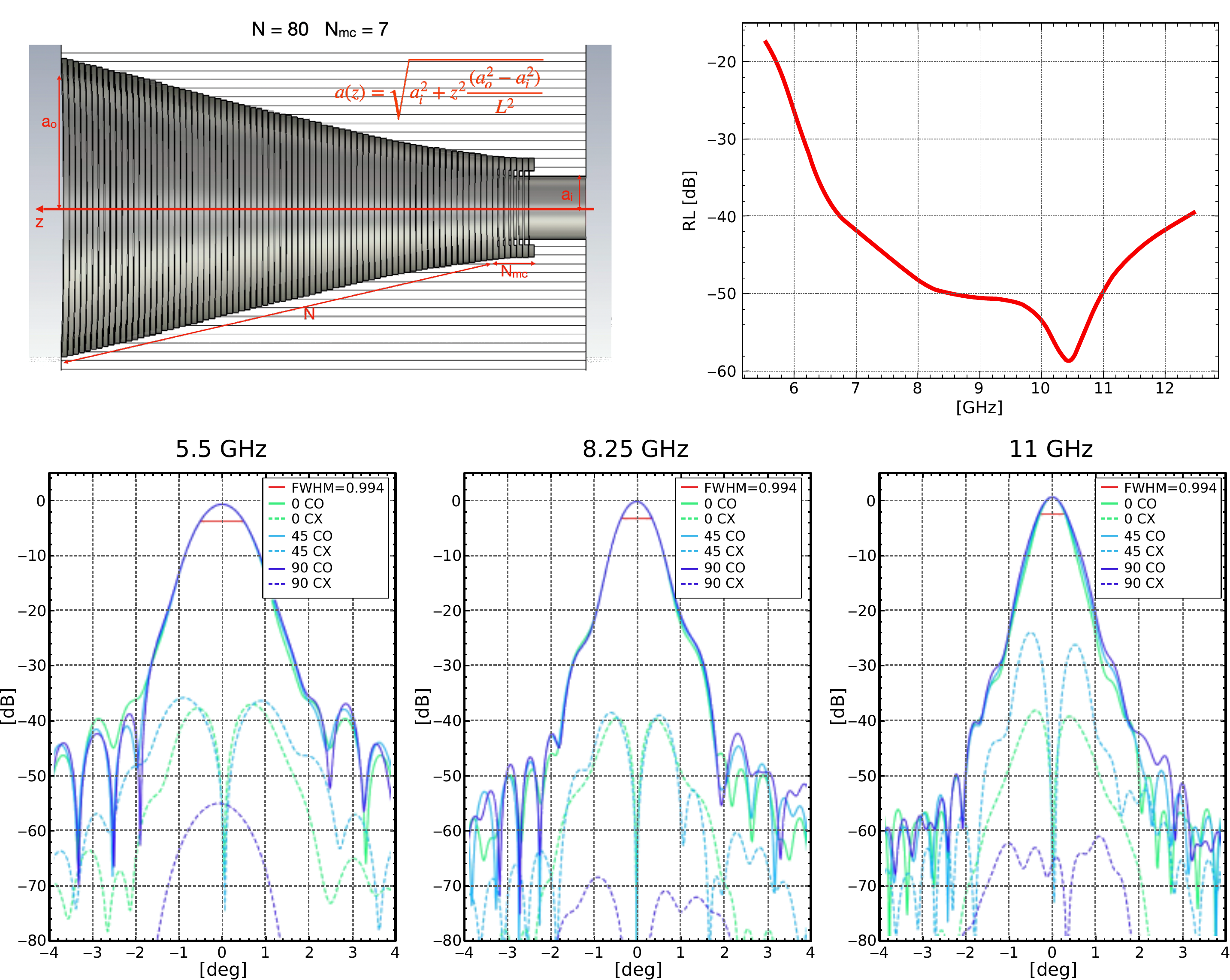}
\end{center}
\caption{\label{fig_feed}
        The ELFS/SA Xband feedhorn. Top-left: the feedhorn design. Top-right: simulated return loss. Bottom panels: the simulated co-polar and cross-polar beams.}
\end{figure*}

The sky signal reflected by the telescope propagates through a room-temperature corrugated feedhorn. Fig.~\ref{fig_feed} shows the feedhorn electromagnetic design with the main dimensions and parameters. We based the design on the ideas presented in \cite{granet2005}, choosing a hyperbolic profile and a ring-loaded mode converter. The corrugation teeth are $4.5$\,mm in the body of the horn and $5$\,mm in the mode converter part, with constant tooth/groove ratio of $0.889$. This configuration allowed us to obtain excellent broadband performance in terms of return loss and cross-polarization performance. The return loss is better than $-20$\,dB and for a large fraction of the band (at frequencies $\gtrsim 6.5$\,GHz) it is less than $-40$\,dB. The beam is highly symmetric, with a maximum cross-polarization of the order of $-40$\,dB, apart from the high edge of the band where it increases to $-30$\,dB. The sidelobes are less than $-50$\,dB across all the band.

\begin{figure*}
\begin{center}
    \includegraphics[width=\textwidth]{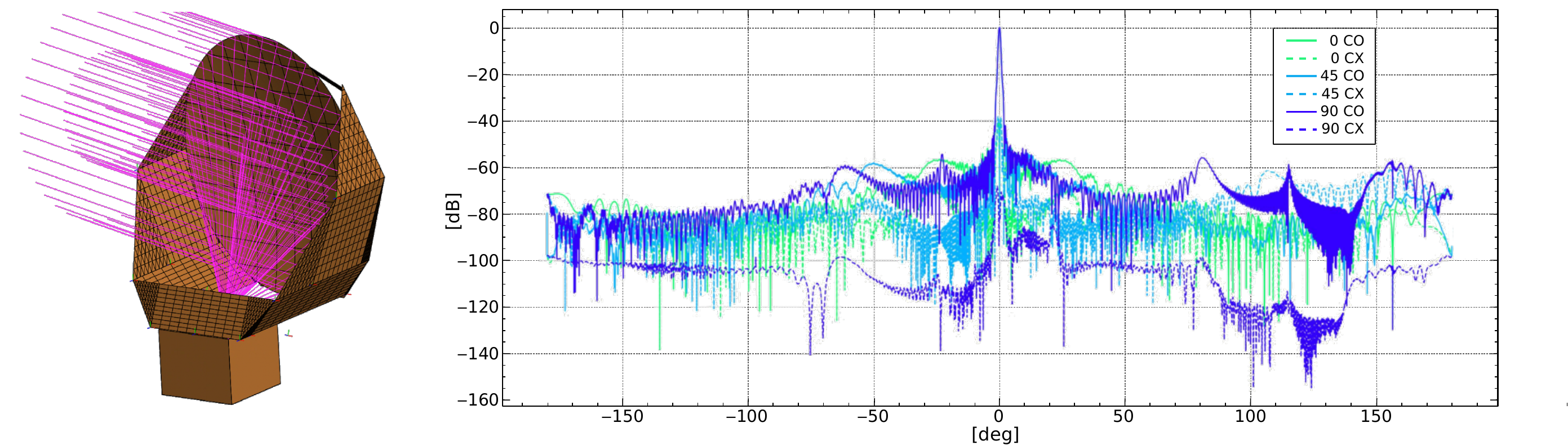}
\end{center}
\caption{\label{fig_feed-telescope}
        The ELFS/SA Xband optical system performance. Left: ray-tracing of the feedhorn at the focus of the SA telescope. Right: simulated beam pattern of the entire optical system.}
\end{figure*}

Fig.~\ref{fig_feed-telescope} shows a physical optics model and full-sky beam simulation of the SA telescope fed by this horn, made using the software package {\tt GRASP}. The simulation was performed neglecting the presence of the absorbing baffling structure which surrounds the primary and secondary mirrors. We see that even in this worst-case scenario most of the far sidelobes are well below $-60$\,dB maximum and the cross-polar response never exceeds $-50$\,dB. In practice the baffling will improve this performance even further.

\begin{figure*}
\begin{center}
    \includegraphics[width=16cm]{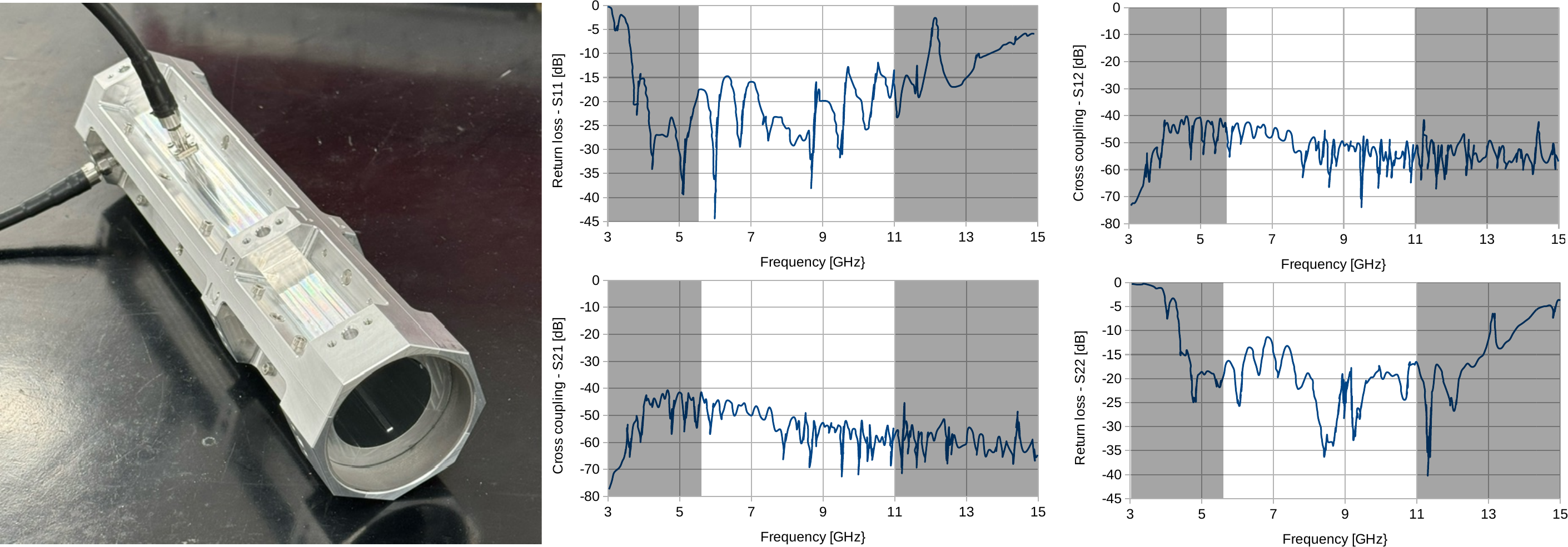}
\end{center}
\caption{\label{fig_omt}
        The ELFS/SA OMT. Left: the fabricated OMT. Right: measured OMT performance. Top-left and bottom-right: return loss at the output ports. Top-right and bottom-left: cross-coupling between the output ports. The white area highlights the working band.}
\end{figure*}

The cold ($4$\,K) OMT splits the signal into two orthogonal linear polarizations, which are then combined by a $90^\circ$ hybrid to produce left- and right-hand circularly polarized signals. The OMT design is based on that used for the Square Kilometre Array (SKA) Band $5$a and $5$b feeds. It is a quad-ridge design, where the ridges for one polarization are maintained at a constant spacing as they pass the coaxial probe and backshort of the other polarization. This configuration minimizes polarization coupling and supports a bandwidth exceeding $2$:$1$ while suppressing excitation of higher-order modes. Fig.~\ref{fig_omt} shows the manufactured OMT and the measured return loss and cross-coupling at the output ports. Within the operating band the return loss is generally better than $-15$\,dB and the cross-coupling is below $-40$\,dB.

Low-noise amplifiers (LNAs), model LNF\_LNC4\_16C from Low Noise Factory, provide the primary gain. Coaxial $30$\,dB couplers in the RF lines before the LNAs enable the injection of a noise signal for calibration purposes. The ability to select in-phase or quadrature versions of the noise signal enables the calibration of $I$, $Q$ and $U$.

The back-end employs an FPGA-based digital unit, sharing the same design as the QUIJOTE-MFI2 instrument \cite{MFI2}. The MFI2 backend, based on the Xilinx ZCU208 Ultrascale board, can simultaneously acquire eight RF channels at a sampling frequency of $5.0$\,GSps, offering a $2.5$\,GHz bandwidth with $1$\,MHz spectral resolution. The back-end divides the full bandwidth into spectral sub-bands, each with a maximum bandwidth of $2.5$\,GHz, which are down-converted to a baseband range of $[0,2.5]$\,GHz using separate local oscillators (LOs). Down-conversion is achieved through the complex output of mixers, enabling two bands to be extracted per LO.

The back-end design integrates a polyphase filter bank and a fast Fourier transform (FFT) to extract spectral information from the digitized time-domain samples produced by the onboard analog-to-digital converters (ADCs). The resulting power spectral density is then integrated over time and averaged. A temporary storage buffer can hold up to 24 hours of raw data, allowing for the identification and removal of undesired interference via an optimized blocking filter. The final stored scientific signal is a spectrally averaged output, with the averaging factor determined by the scientific objectives and available storage capacity.

\section{Simulations} \label{sec:simulations}

In this section, we describe the pipeline used to generate the signal simulations for each instrument configuration considered in the analysis. We consider the addition of data from the low-frequency instruments XBand (phase 1 of ELFS/SA) and MFI2 (phase 2 of ELFS/SA) to that from a representative future CMB experiment, hereafter referred to as ngCMB. For the purposes of our analysis we take the baseline ngCMB configuration to be similar to that of the nominal SO experiment with specifications as outlined in \cite{so_forecast_2023} but with enhanced capabilities and observation time as might plausibly be achieved with upcoming additions to SO (e.g. SO-Japan and SO-UK) or with the prospects of Stage-4 CMB experiments. We assume that all these instruments are capable of observing the same celestial region from a site in the Atacama Desert. Furthermore, to ensure that our results are directly comparable to the latest SO forecast paper \cite{so_forecast_2023}, we have generated our maps using the HEALPix pixelation scheme \cite{HEALPix} with a resolution parameter set to $N_{\rm side} = 512$. In addition, we have adopted delta-like frequency bandpasses, aligning our methodology with that employed in their study.

We also describe the different components included in the simulations. First, we review the cosmological and astrophysical signals that are common to all instruments. These are detailed in Section~\ref{subsec:sky_signal}. In Section~\ref{subsec:noise_simus}, we present the formalism and specifications used to simulate the noise for each instrument.

\begin{table}
    \caption{Sky components included in the simulations.}
    \label{tab:components_sims}
    \centering
       \begin{tabular}{ccl}
    \toprule
    Component   & code & Characteristics \\
    \midrule
    CMB & \texttt{cL} & Lensed CMB ($a_L=1$). \\
     \midrule
     \multirow{4}{*}{Synchrotron} & \texttt{s5} & PySM \texttt{s5} model (power law). \\
     & \texttt{s7} & PySM \texttt{s7} model (power law with curvature)\\
     & \texttt{sC2} & Custom model with random Gaussian curvature.\\
     & \texttt{sC} & Custom model scaling the \texttt{s7} $c_s$ template.\\
     \midrule
     Thermal dust & \texttt{d10} & PySM \texttt{d10} model (modified black-body). \\
    \toprule
    \end{tabular}        
\end{table}

\subsection{Sky signal}
\label{subsec:sky_signal}

The polarized microwave sky is modelled as consisting of the CMB, and foreground emission from synchrotron radiation at low frequencies and thermal dust emission at high frequencies. In Section~\ref{subsubsec:cmb}we describe the simulation of the CMB signal, and in Section~\ref{subsubsec:fg} the primary diffuse foregrounds, namely synchrotron and thermal dust. In Section~\ref{subsubsec:other} we discuss other possible effects such as Faraday rotation and the presence of polarized Anomalous Microwave Emission (AME).  A summary of the different components simulated and the models used can be found in Table~\ref{tab:components_sims}. To account for the instrumental beam, both CMB and foreground maps are convolved with a Gaussian beam. The corresponding FWHMs are reported in Table~\ref{tab:noise_properties}.

\subsubsection{CMB}
\label{subsubsec:cmb}

We conduct a series of $50$ CMB simulations, treating them as Gaussian random realizations derived from power spectra\footnote{The power spectra are computed using the \texttt{CAMB} software \cite{camb}.} which are compatible with the best-fit cosmological parameters obtained from analysis of the \textit{Planck} data \cite{planck_cosmo_par}. In our modelling, we adopt the assumption of no tensor fluctuations ($r=0$), thereby attributing any potential detection of $r$ to residual foreground effects.

\subsubsection{Foregrounds}
\label{subsubsec:fg}
Our analysis encompasses several models, which are detailed below:
\begin{itemize}
    \item \textbf{Synchrotron.} We use four different synchrotron models. The first and second are the \texttt{s5} and \texttt{s7} models respectively from  \texttt{PySM}, which assume a simple power law and a power law with negative curvature with a pivot frequency of $23$\,GHz.
 For the third and fourth models, we generate two different curvature map templates which are compatible with current constraints from Southern Hemisphere measurements. Specifically, the curvature templates are compatible with the value of $c_s = 0.04 \pm 0.1$ found in \cite{synchrotron_spass}. The resulting template maps are shown in column one of Figure~\ref{fig:c_s_maps}. The \texttt{sC2} template (middle-left in Figure~\ref{fig:c_s_maps}), is generated by drawing a random Gaussian realization of the distribution $\mathcal{N}\left(0.04, 0.1\right)$ at $N_{\rm side}=16$. This template is then upgraded to a resolution of $N_{\rm side} = 2048$ and smoothed with a Gaussian beam of $5^{\circ}$. The \texttt{sC} template (bottom-left in Figure~\ref{fig:c_s_maps}) is obtained by renormalizing the $c_s$ template of the \texttt{PySM} \texttt{s7} model to a Gaussian distribution with a mean of $0.04$ and a standard deviation of $0.1$. This template shows an anisotropic distribution of curvatures similar to the \texttt{s7} template, where the values smaller than the template average are located in regions with high synchrotron signal-to-noise, and vice versa.

Unlike the \texttt{s7} model, the \texttt{sC2} and \texttt{sC} models contain pixels with positive curvature at a pivot frequency of $23$\,GHz. Positive curvature results in a flatter synchrotron spectrum and hence a higher synchrotron component at CMB frequencies (i.e $\ge$ $90$\,GHz). These models therefore represent a more pessimistic scenario for synchrotron contamination compared to the \texttt{s7} model. It is important to note that our current understanding of the synchrotron SED is limited; the models studied here remain consistent with existing constraints. This is why we explore various synchrotron scenarios in this work.

\item \textbf{Thermal dust.} Thermal dust radiation originates from dust grains present in the interstellar medium. The grains absorb ultraviolet light and re-emit as a grey body. In general, these dust grains are not perfectly spherical and typically have their minor axis aligned with the direction of the local magnetic field, which results in polarized thermal dust emission. 

While more complex models might be necessary to characterize dust emission \cite{dust_mcbride,dust_ritacco}, this paper focuses primarily on synchrotron emission. Thus, we adopt a single dust model (\texttt{PySM} \texttt{d10}), which assumes that dust behaves as a modified black-body, with a power-law dependence of emissivity on frequency.

\end{itemize}

\subsubsection{Other components/effects}
\label{subsubsec:other}

The polarized nature of the AME remains a subject of ongoing investigation, primarily because its exact nature remains uncertain \cite{ame_review}. Several proposed models include spinning dust particles \cite{Haimoud2013}, magnetic dipole emission \cite{Draine1999}, and more recently, the suggestion of spinning nano-diamonds \cite{Greaves2018}. These models generally predict a polarization fraction for AME emission of less than $5$\%.

In terms of data analysis, studies that focus on compact regions have not yielded conclusive evidence of polarization. The most rigorous constraints on the polarization fraction, denoted as $\Pi$, were provided by \cite{Genova2017}, indicating $\Pi < 0.22\%$ at $41$\,GHz. Due to this absence of empirical support, we have opted not to consider the AME component in our work.

Another issue intrinsic to the polarization signal is the Faraday rotation effect, i.e., the rotation of the plane of polarization that occurs when light passes through the interstellar medium in the presence of a magnetic field. Due to the finite size of the beam, this effect can lead to beam depolarization, as the measured signal is an average of the emission from various directions within the beam with different rotation angles.   

The Faraday rotation angle scales with the square of the wavelength, hence its effects are more significant at low frequencies. To estimate the magnitude of this effect at our frequencies of interest, we calculate the rotation angles induced by the Faraday rotation using the model from \cite{Faraday_Rotation}. From these estimates of the rotation angle\footnote{The maps are shown in Appendix~\ref{sec:appendix_fr}. 
}, we find that most pixels are rotated by $\phi<6$\,deg. This suggests that this effect can be safely neglected in most cases. Notably, the largest deviations occur close to the Galactic plane. In the event that masking becomes necessary due to an imperfect understanding of this effect, a substantial portion of the sky remains available, preserving the most sensitive data.

\subsection{Noise simulations}
    \label{subsec:noise_simus}

\begin{table}
    \centering
    \caption{Instrument and noise specifications. The columns include the central frequency ($\nu$) in GHz, the beam full-width at half-maximum (FWHM) in arcmin, the channel sensitivity ($N_{\rm white}$) in $\mu$K-arcmin, knee frequency in power spectrum space ($\ell_{\rm knee}$), and the correlated noise slope ($\alpha_{\rm knee}$).}
    \begin{tabular}{cccccc}
    \toprule
    Experiment   &  $\nu$ & FWHM  &  N$_{\rm white}$  & $\ell_{\rm{knee}}$ & $\alpha_{\rm{knee}}$\\
    \midrule
    \multirow{6}{*}{ngCMB} & 27 & 91    & 11.0    & 15   & -2.4 \\
    & 39    & 63    & 7.33    & 15    & -2.4 \\
    & 93    & 30    & 1.25   & 25    & -2.5 \\
    & 145   & 17    & 1.40   & 25    & -3   \\
    & 225   & 11    & 3.71   & 35    & -3   \\
    & 280   & 9     & 9.44     & 40   & -3   \\
    \midrule
    \multirow{6}{*}{Xband} & 6.3 & 46.6  & 539    & 15    & -2.4 \\
    & 7   &  42.2     & 512    & 15    & -2.4  \\
    & 7.7   & 38.1    & 487    & 15    & -2.4 \\
    & 8.6   & 34.4    & 465   & 15    & -2.4 \\
    & 9.5   & 31.1    & 443   & 15    & -2.4 \\
    & 10.5  & 28.1    & 423   & 15    & -2.4 \\
    \midrule
    \multirow{5}{*}{MFI2} & 10.5 & 33.7  & 245    & 15    & -2.4 \\
    & 12.9   &  27.4   & 228    & 15    & -2.4  \\
    & 14.3   & 24.8    & 206    & 15    & -2.4 \\
    & 15.9   & 22.2    & 236   & 15    & -2.4 \\
    & 18.4   & 19.2    & 203   & 15    & -2.4 \\
    \toprule
    \end{tabular}
    \label{tab:noise_properties}
\end{table}

In this section, we outline our procedure for generating noise simulations for the three instruments under consideration in our analysis: ngCMB, Xband (phase 1 of ELFS/SA) and MFI2 (phase 2 of ELFS/SA).

We generate $50$ noise simulations for each instrument, following the same methodology as detailed in \cite{SO,so_forecast_2023}. Initially, we model the noise power spectra for each band of each instrument as the sum of two noise sources, which are:
\begin{equation}
    N_{\ell} = N_{\rm white}\left[ 1 + \left(\dfrac{\ell}{\ell_{\rm knee}}\right)^{\alpha_{\rm knee}} \right] \, ,
    \label{eq:red_noise_ps}
\end{equation}
a white noise component represented by a constant $N_{\rm white}$, alongside a $1/f$ component\footnote{Analogously to \cite{SO}, we assume that the power of the $1/f$ component equals that of the white noise component.} modeled as a power law at the power spectrum level. This $1/f$ component is characterized by an exponent denoted as $\alpha_{\rm knee}$ and an elbow multipole at $\ell_{\rm knee}$. While the white noise accounts for the instrument sensitivity, the $1/f$ component models the correlated noise that arises from the atmosphere and electronic noise.

To generate a noise simulation for a specific band, we draw a random Gaussian realization of the noise power spectra from Equation~\ref{eq:red_noise_ps} and then rescale the map by $\sqrt{N_{\rm hits}}$\footnote{We employ the simulated map of hit counts presented in Figure 8 of \cite{SO}.} to account for the inhomogeneous coverage.

Noise specifications for the different bands are documented in Table~\ref{tab:noise_properties}, covering ngCMB, X-band, and MFI2. The ngCMB specifications are estimated based on the nominal Simons Observatory (SO) values, particularly the optimistic goal case outlined in \cite{so_forecast_2023}. These sensitivities are then rescaled to account for additional contributions from SO:JP and SO:UK\footnote{\url{https://simonsobservatory.org/fact-sheets/}}, as well as an extended observational period:

\begin{itemize}
\item ngCMB includes two additional medium-frequency (MF) SATs, as per the planned addition of SO:UK, dedicated to measuring the 93 GHz and 145 GHz bands.
\item To model the expected SO:JP contribution, an additional SAT is allocated exclusively for low-frequency (LF) measurements, covering the 27 GHz and 39 GHz bands throughout the entire campaign. Unlike the nominal SO case, where one SO-SAT was used for LF measurements for only a single year, this dedicated LF SAT will operate for the full duration.
\item Instead of assuming a 5-year observation period, we consider ngCMB to be operational until 2035. As a result, the nominal SO-SAT-like telescopes will operate for 11 years, while the SO:UK-like and SO:JP-like SATs will be in operation for 9 years.
\end{itemize}

Based on these assumptions, the scaling factors for LF, MF, and HF\footnote{High frequency, corresponding to the $230$ and $280$\,GHz bands.} SATs are $\sqrt{1/9}$, $\sqrt{5/(9+11)} = 1/2$\footnote{We do not account for the additional year of MF measurements since the SO:JP SAT will now conduct the LF measurements. The scale factor in this case is approximately $\sqrt{9/(18+22)}\sim 0.47$. This minor change does not significantly impact our study, as the MF frequencies have limited relevance for characterizing synchrotron emission.}, and $\sqrt{5/11}$, respectively. 

The Xband sensitivity is calculated as a single-pixel system of the PolarBear experiment assumimg a sky fraction of $f_{\rm sky}=0.1$, equivalent to the SO-SAT sky patch, and an observing time of $6$ years with $25$\% observation efficiency. The MFI2 sensitivity is likewise obtained by scaling its expected sensitivity \cite{MFI2} to that sky patch for an observing time of $5$ years with $25$\% observation efficiency. For the correlated noise, we utilize the $\ell_{\rm knee}$ and $\alpha_{\rm knee}$ values from SO's lowest frequency band \cite{so_forecast_2023} for both Xband and MFI2.

\begin{figure}
    \centering
    \includegraphics[width=.75\linewidth]{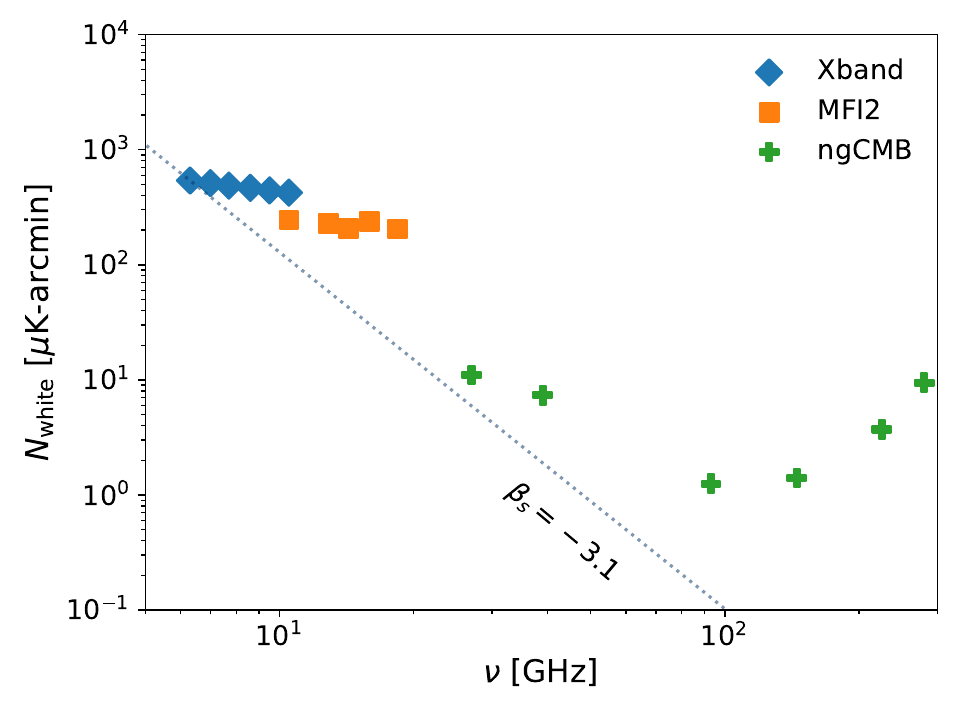}
    \caption{Sensitivities as a function of frequency. The dotted lines shows the extrapolated sensitivity following a power law with $-3.1$ exponent.}
    \label{fig:nwhite_per_exp}
\end{figure}

Figure~\ref{fig:nwhite_per_exp} shows the sensitivity of the bands as a function of their frequency. The dotted lines illustrate the Xband and MFI2 equivalent sensitivity at other frequencies, extrapolated using a power law with an exponent of $-3.1$. Notably, Xband exhibits a more favorable relative sensitivity than the lowest ngCMB band.

\begin{table*}
\centering
\scriptsize
\abovedisplayskip=0pt
\belowdisplayskip=0pt
\caption{Parametric models and prior information used in the analysis. Equations are given in antenna units. $B(\nu,T)$ is the Planck function. $x=\frac{h\nu}{k_B T_d}$, $h$ and $k_B$ are Planck and Boltzmann constants respectively. }
\label{tab:priors_par_models}
\hspace*{-.9cm}
\begin{tabular}{l l m{6.8cm} @{\hskip 20pt}>{\raggedright\arraybackslash}m{3.5cm}}
\toprule
Component & Model & Equation  & Priors \\
\midrule
CMB & Black-body & {\begin{flalign}
\begin{pmatrix}
        Q_{\mathrm{cmb}} \\
        U_{\mathrm{cmb}}
    \end{pmatrix}
    =
    \begin{pmatrix}
        \mathrm{cmb}^Q \\
        \mathrm{cmb}^U
    \end{pmatrix}
    \dfrac{x^2e^x}{(e^x-1)^2} &&
\label{eq:cmb_model}\end{flalign}} &   \\ \hdashline
\multirow{5}{*}{Synchrotron} & Power law & {\begin{flalign}
\begin{pmatrix}
        Q_s \\
        U_s
    \end{pmatrix}_{\nu}
    =
    \begin{pmatrix}
        a_s^Q \\
        a_s^U
    \end{pmatrix}
    \left(\frac{\nu}{30 \,\mathrm{GHz}}\right)^{\beta_s}  &&
\label{eq:synchrotron_model}\end{flalign}} &  $\beta_s  \in \mathcal{N}(-3.1, 0.1)$\\ 
 & Power law with curvature & {\begin{flalign}
\begin{pmatrix}
        Q_s \\
        U_s
    \end{pmatrix}
    =
    \begin{pmatrix}
        a_s^Q \\
        a_s^U
    \end{pmatrix}
    \left(\frac{\nu}{30\,\mathrm{GHz}}\right)^{\beta_s + c_s \log\left(\frac{\nu}{23 \mathrm{GHz}}\right)}  &&
\label{eq:synchrotron_model_curv}\end{flalign}} &
\[
\begin{array}[t]{@{}l@{}}
\beta_s \in \mathcal{N}(-3.1, 0.1) \phantom{kdkdkddkd} \\
c_s \in \mathcal{N}(0, 0.1)
\end{array}
\]\\ \hdashline
Thermal dust & Modified black-body & {\begin{flalign}
\begin{pmatrix}
        Q_d \\
        U_d
    \end{pmatrix}
    =
    \begin{pmatrix}
        a_d^Q \\
        a_d^{U}
    \end{pmatrix}
    \left(\frac{\nu}{353\,\mathrm{GHz}}\right)^{\beta_d-2}\frac{B(\nu, T)}{B(\nu, T_d)}&&
\label{eq:dust_model}\end{flalign}} &   {\begin{align*}
\beta_d & \in \mathcal{N}(1.54, 0.1) \cup \mathcal{U}(1,2.4) \\
T_d  &\in         \mathcal{N}(21, 3) \cup \mathcal{U}(10, 40)  \end{align*}}  \\
\hline\end{tabular}
\end{table*}

\section{Methodology} \label{sec:methodology}

Here, we outline the steps taken to calculate the tensor-to-scalar ratio, which involves two main phases. First, we perform a component separation analysis to obtain a clean CMB map, as explained in Section~\ref{subsec:ml_method}. Next, we determine the cosmological parameter $r$ from the cleaned CMB power spectrum  by sampling the cosmological likelihood, as described in Section~\ref{subsec:cosmo_likelihood}.

\subsection{Component Separation}
\label{subsec:ml_method}

To perform the component separation analysis we apply the \texttt{B-SeCRET} methodology \cite{de2020detection}. \texttt{B-SeCRET} is a parametric pixel-based maximum-likelihood method, which relies on an Affine-Invariant Markov Chain Monte Carlo Ensemble sampler (\texttt{emcee}\footnote{\url{https://emcee.readthedocs.io/en/stable/}}) to draw samples from a posterior distribution\footnote{The best-fit parameters' map are evaluated as the median of each marginalized parameter pixel posterior probability.} \cite{foreman2013emcee}. Preceding the component separation analyses, the frequency maps are convolved appropriately to have a uniform beam characterized by a FWHM of $91$\,arcmin, and downgraded to the same resolution, given by the HEALPix \cite{HEALPix} parameter $N_{\rm side} = 64$. The pipeline followed to downgrade the maps is described in \cite{synch_mfi}.

\texttt{B-SeCRET} applies Bayesian inference to determine the best-fit model parameters given some prior information. In Bayesian statistics, the probability of the model parameters $\myset{\theta}\polarization_{p}$ given the signal data $\myvector{d}\polarization_{p}$ at the pixel $p$ is proportional to the probability of the $\myvector{d}\polarization_{p}$ given $\myset{\theta}\polarization_{p}$ times the  probability of $\myset{\theta}\polarization_{p}$, i.e.,
\begin{equation}
    \mathcal{P}(\myset{\theta}\polarization_{p}|\myvector{d}\polarization_{p}) \propto \mathcal{P}(\myvector{d}\polarization_{p}|\myset{\theta}\polarization_{p}) \mathcal{P}(\myset{\theta}\polarization_{p}) \, .
    \label{eq:posterior}
\end{equation}
$\mathcal{P}(\myset{\theta}\polarization_{p})$ is commonly known as the prior information, whereas $\mathcal{P}(\myvector{d}\polarization_{p}|\myset{\theta}\polarization_{p})$ is usually referred to as the likelihood. Assuming Gaussian noise, the likelihood of the data can be expressed as 
\begin{equation}
    \mathcal{P}(\myvector{d}\polarization_{p}|\myset{\theta}\polarization_{p}) = \dfrac{\exp\left(-\dfrac{1}{2}\left(\myvector{d}\polarization_{p}-\myvector{S}_{p}\polarization\right)^{T}\mymatrix{C}^{-1}\left(\myvector{d}\polarization_{p}-\myvector{S}_{p}\polarization\right)\right)}{\sqrt{(2\pi)^{N}\det (\mymatrix{C})}} \, ,
    \label{eq:likelihood}
\end{equation}
where $\mymatrix{C}$ is the covariance matrix, $N =2N_{\nu}$ is the number of elements in the $\myvector{d}_{p}$ array, $N_{\nu}$ is the number of frequency channels, and $\myvector{S}_{p}$ is the following parametric model 

\begin{equation}
    \myvector{S}_p = 
    \begin{pmatrix}
        \myvector{Q}_{S} \\
        \myvector{U}_{S}
    \end{pmatrix}
	=
	\begin{pmatrix}
        \myvector{Q}_{\rm cmb} \\
        \myvector{U}_{\rm cmb}
    \end{pmatrix}
    +
    \begin{pmatrix}
        \myvector{Q}_{s} \\
        \myvector{U}_{s}
    \end{pmatrix}
    +
    \begin{pmatrix}
        \myvector{Q}_{d} \\
        \myvector{U}_{d}
    \end{pmatrix}
    \, ,
\label{eq:sky_model}
\end{equation}
where $\myvector{X}_{\rm cmb}$, $\myvector{X}_{s}$, and $\myvector{X}_{d}$ are the CMB, synchrotron, and thermal dust models evaluated at the instrument frequencies $\myvector{\bm{\nu}}$ respectively ($X \in \{Q,U\}$). The different parametric models considered for each component are summarized in Table~\ref{tab:priors_par_models}. 

As part of our methodology, we incorporate Gaussian priors on the component SED spectral parameters. These specific priors, along with their respective details, can be found in Table~\ref{tab:priors_par_models}.  The inclusion of priors is especially crucial in scenarios with high parameter degeneracy. For instance, in the context of the power law with curvature model, the parameters $\beta_s$ and $c_s$ exhibit a high degree of degeneracy. In cases where the signal-to-noise ratio is insufficient, many combinations of $\beta_s$ and $c_s$ can yield a close match to the effective exponent at the lowest frequency $\beta_{\rm eff}\left(\nu_{\rm min}\right) = \beta_s + c_s\log\left(\nu_{\rm min}/ 23\,\mathrm{GHz}\right)$ considered in the analysis. This degeneracy can be visualized as a line in the ($\beta_s$, $c_s$) plane, with a slope determined by $-\log \left(\nu_{\rm  min}/23,\mathrm{GHz}\right)$.

Instead of conducting a comprehensive sampling of the full posterior distribution, we split the model parameters into two categories: amplitudes and spectral parameters. We then proceed to sample the conditional probability distributions separately. This strategy effectively diminishes the dimensionality of the problem, meaning we have a smaller space to explore, ultimately aiding in achieving convergence. We continue this process iteratively, sampling the conditional probability distributions until we reach a state of convergence.

\begin{figure*}
    \centering
    \includegraphics[width=\linewidth, trim={2.5cm 0cm 3.25cm 1cm}, clip]{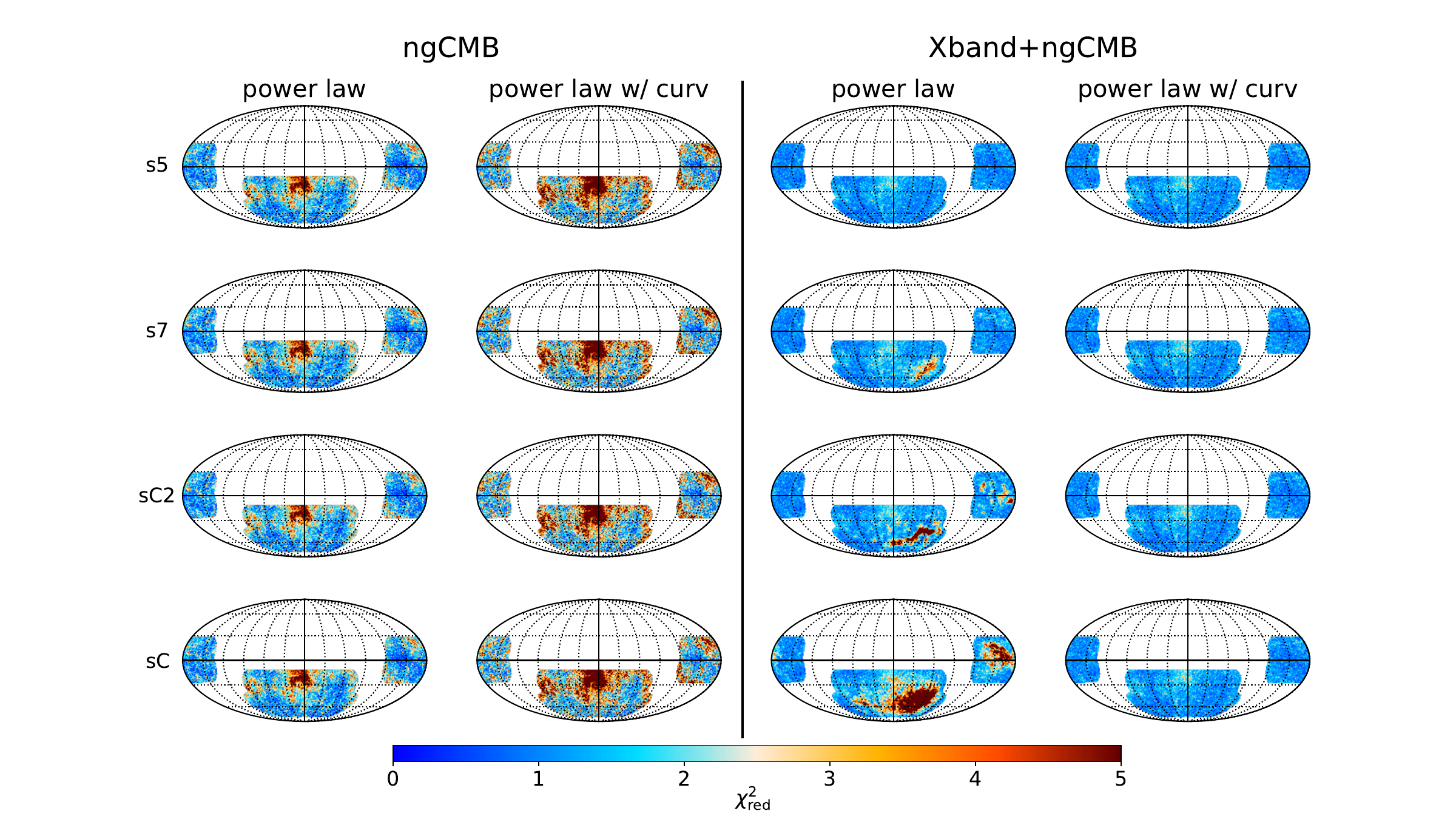}
    \caption{$\chi^2_{\rm red}$ maps obtained using only ngCMB (first and second columns) and Xband+ngCMB (third and fourth column). Rows show the results obtained when the synchrotron is simulated using \texttt{s5}, \texttt{s7}, \texttt{sC2}, \texttt{sC} from top to bottom. The odd (even) columns show the results obtained when the synchrotron is modeled in the component separation analysis using either a power law (power law with curvature). }
    \label{fig:chi2}
\end{figure*}

    \subsection{Cosmological Likelihood} \label{subsec:cosmo_likelihood}

We calculate the power spectra from the cleaned CMB map using a python implementation of a pure pseudo-$C_{\ell}$ algorithm \texttt{NaMaster}\footnote{\url{https://namaster.readthedocs.io/en/latest/}.} \cite{namaster}. We apodize the mask using a square cosine function with a kernel of $22$\,deg, resulting in an effective sky fraction of approximately $f_{\rm sky} \sim 10$\%, and use a uniform binning scheme. Each bin contains $5$-$\ell$ multipoles that go from $\ell \in [30,128]$. The multipoles with a given bin $q$ are weighted with $w_{\ell} = (\ell +1/2)/ \sum_{\ell \in q} (\ell+1/2)$ so that larger multipoles have more weight.

We then calculate the tensor-to-scalar ratio by sampling\footnote{We use the Python library \texttt{emcee} to sample the likelihood.} the following likelihood:
\begin{equation}
    \log \mathcal{L} \propto -\dfrac{1}{2}\left(\widehat{\mathbf{C}}_{q} - \mathbf{C}_{q}\right)^T\mathbf{M}\left(\widehat{\mathbf{C}}_{q} - \mathbf{C}_{q}\right) \, ,
    \label{eq:gauss_likelihood}
\end{equation}
where $\widehat{\mathbf{C}}_{q}$ is the cleaned CMB power spectrum, $\mathbf{M}$ is the Gaussian covariance matrix among $q$ bins obtained from \texttt{Namaster}, and $\mathbf{C}_{q}$ is the theoretical model:
\begin{equation}
    \mathbf{C}_{q}(r, a_L) = r\mathbf{C}^{\mathrm{PGW} (r=1)}_{q} + a_L \mathbf{L}_{q} + \mathbf{N}_q\, ,
\end{equation}
where $\mathbf{C}^{\mathrm{PGW} (r=1)}_{q}$ is the primordial BB power spectrum, $\mathbf{L}_{q}$ is the lensing BB power spectrum, and $\mathbf{N}_q$ is a model of the noise power spectrum estimated from simulations. 
\section{Results} \label{sec:results}

\begin{figure}
    \centering
    \includegraphics[width=0.49\linewidth, trim={2cm 9cm 2cm 0.5cm}, clip]{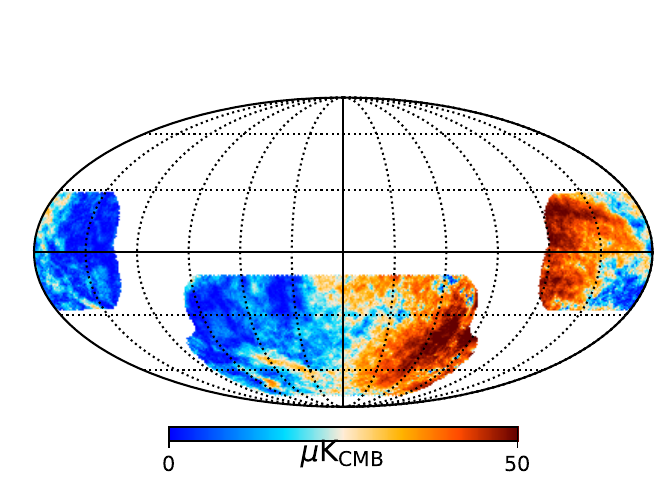}
    \includegraphics[width=0.49\linewidth, trim={2cm 1.55cm 2cm 9.5cm}, clip]{new_figures/pol_amp_s2n_synch_s5_27GHz.pdf}
    \caption{
        \textit{Left:} Polarized synchrotron amplitude intensity at $27$\,GHz.
        \textit{Right:} Synchrotron signal-to-noise with ngCMB sensitivity at $27$\,GHz. Maps are in equatorial coordinates.
    }
    \label{fig:pol_int_synch_amp_27GHz}
\end{figure}

\begin{figure*}
    \centering
    \includegraphics[width=\linewidth, trim={3cm 0cm 3cm 1cm}, clip]{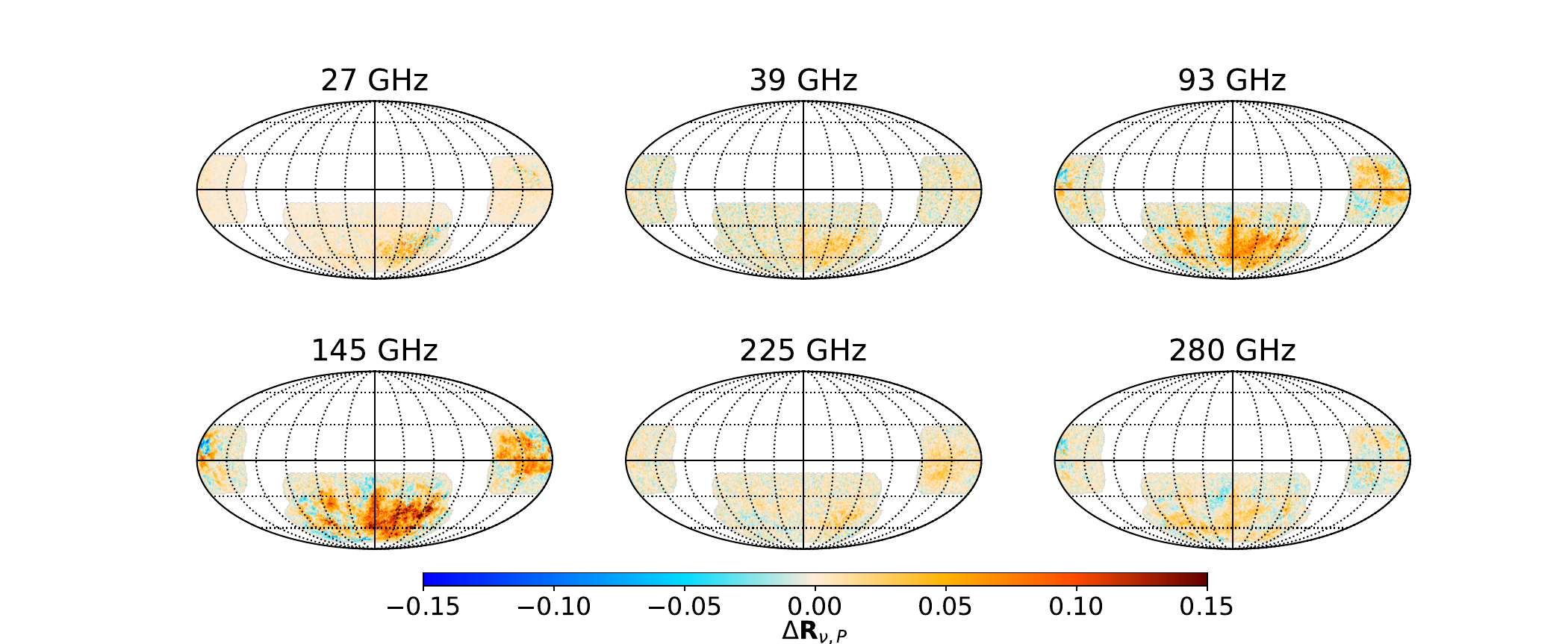}
    \caption{Difference maps between the residual maps obtained when the synchrotron is simulated with \texttt{sC} versus \texttt{s5} for ngCMB alone.}
    \label{fig:diff_residuals}
\end{figure*}

\begin{figure*}
    \centering
    \includegraphics[width=\linewidth, trim={2.5cm 1cm 3.25cm 2cm}, clip]{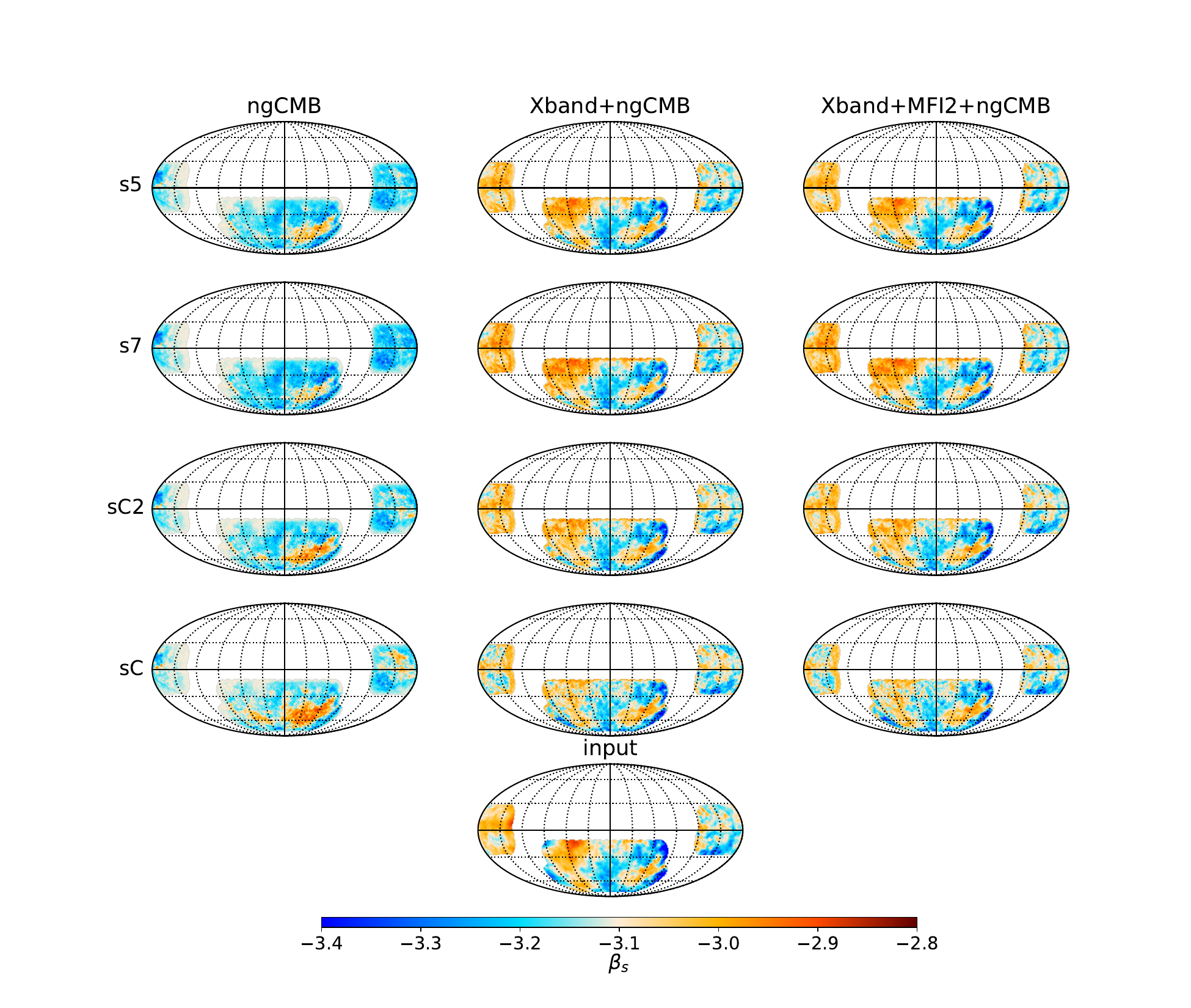}
    \caption{Recovered $\beta_s$ maps for ngCMB (left column), Xband+ngCMB (middle column), and Xband+MFI2+ngCMB (right column). Rows $1$–$4$ display the results for the \texttt{s5}, \texttt{s7}, \texttt{sC2}, and \texttt{sC} skies, respectively. Row $5$ shows the input $\beta_s$ map, downgraded to $N_{\rm side} = 64$.}
    \label{fig:beta_s}
\end{figure*}

This section is organized as follows: Section~\ref{subsec:synch_sed} discusses the accuracy of characterizing the Synchrotron's SED. In Section~\ref{subsec:cosmology}, we evaluate the ability to recover the tensor-to-scalar ratio. For each analysis, we compare the achievable results using the ngCMB data without supplementary low-frequency information and assess the impact of incorporating ELFS/SA data.

\subsection{Synchrotron Characterization}
\label{subsec:synch_sed}

This subsection examines the information that can be obtained about the synchrotron emission from ngCMB data alone, as well as from the combination of ngCMB data with ELFS/SA. The analysis focuses on two main aspects: evaluating the synchrotron SED using goodness-of-fit estimators (Section~\ref{subsubsec:goodness-of-fit}) and assessing the accuracy of spectral parameter recovery (Section~\ref{subsubsec:spec_par}).

\subsubsection{SED}
\label{subsubsec:goodness-of-fit}

Parametric methods allow us to assess goodness-of-fit and conduct model selection, i.e., enabling us to determine the most suitable model for the observed data. Here, to test which model provides a better goodness-of-fit, we use the reduced $\chi^2$ estimator:
\begin{equation}
    \chi^2_{{\mathrm{red},p}} = \dfrac{1}{N_{\rm{dof}}}\sum\limits_{i\in \{Q,U\}}(\myvector{d}_{p,i}-\myvector{S}_{p,i})\mymatrix{C}^{-1}_{p}(\myvector{d}_{p,i}-\myvector{S}_{p,i})\, ,
    \label{eq:chi2_red_regions}
\end{equation}
where $\mymatrix{C}_{p}$ is the noise covariance matrix at pixel $p$\footnote{In this work, $\mymatrix{C}_{p}$ is a diagonal matrix with diagonal elements $\sigma_{\nu,p}^2$, which are derived by calculating the standard deviation $\sigma_{\nu,p}$ from 500 noise simulations that have been downgraded using the same pipeline applied to downgrade the data.}, ${N_{\rm{dof}}} = (2N_{\nu}-N_{\theta})$, $N_{\nu}$ is the number of frequency channels, and $N_{\theta}$ is the number of model parameters. 

Fig.~\ref{fig:chi2} shows the reduced $\chi^2$ maps obtained using either ngCMB or Xband+ngCMB information for the four different synchrotron scenarios considered. These maps are organized in a 4x2 matrix format, with the rows representing the four considered simulated skies, namely, \texttt{s5}, \texttt{s7}, \texttt{sC2},  and \texttt{sC}, while the columns depict the chosen synchrotron models, i.e., the power-law and the power-law with curvature.

Upon comparing the maps on the first column, it becomes evident that ngCMB alone lacks the discerning power necessary to distinguish between the simulated skies, i.e., the $\chi_{\rm red}^2$ maps obtained are very similar independently of the synchrotron simulation. Furthermore, the introduction of an additional parameter, such as curvature, is consistently penalized, as ngCMB's information does not provide sufficient constraining power.

On the other hand, when we add the Xband, we are now able to observe differences in recovered $\chi^2_{\rm red}$ maps among different simulated skies when we use a power law to fit the synchrotron emission (third column). Since the power law is not the correct model for the  \texttt{s7}, \texttt{sC2},  and \texttt{sC} skies, there are regions with large $\chi^2_{\rm red}$ values, especially where the synchrotron signal-to-noise is high, see Fig.~\ref{fig:pol_int_synch_amp_27GHz}, and the $c_s$ parameter deviates significantly from zero, i.e., the sky is less compatible with a power law.

Unlike with ngCMB, we observe that when we add the curvature parameter we are able to recover better $\chi^2_{\rm red}$ maps for \texttt{s7}, \texttt{sC2},  and \texttt{sC} skies (fourth column). We are now able to put constraints on $c_s$ due to the additional information coming from the Xband. Moreover, in the case of \texttt{s5}, the $\chi^2_{\rm red}$ assuming a power law is slightly better than with a power law with curvature since the model former has less parameters.

Normalized residual maps for individual frequency channels can also be utilized:
\begin{equation}
    \mathbf{R}_{\nu, i} = \dfrac{\myvector{d}_{i}-\myvector{S}_{i}}{\bm{\sigma}_{\nu}} \, ,
    \label{eq:res_maps}
\end{equation}
to identify which channels are most affected by incorrect modeling of the sky.  Fig.~\ref{fig:diff_residuals} shows the difference between the normalized polarized intensity residual maps:
\begin{equation}
    \mathbf{R}_{\nu, P} = \sqrt{\mathbf{R}_{\nu, Q}^2 + \mathbf{R}_{\nu, U}^2  } \, ,
\end{equation} 
obtained when the synchrotron is simulated using \texttt{sC} versus \texttt{s5} in the case of ngCMB data alone. 

The results indicate that the medium-frequency channels, which are crucial for CMB reconstruction, are most affected by inaccurate sky modeling. This issue arises because the recovered value of $\beta_s$ is close to the effective value of the spectral index for the power-law-with-curvature model:
\begin{equation}
\beta_{\rm eff}\left(\nu\right) = \beta_s + c_s\log\left(\dfrac{\nu}{23,\mathrm{GHz}}\right) \, ,
\label{eq:beta_eff}
\end{equation}
at the lowest frequency included in the analysis, where the synchrotron signal is spectrally the most dominant. While this yields accurate results at low frequencies, it introduces larger errors in medium-frequency channels, where synchrotron contributions remain significant.

The largest discrepancies occur in regions with both high synchrotron S/N and positive curvature. Positive curvature causes a slower decline in synchrotron emission with increasing frequency, aggravating errors in these regions.

\begin{figure*}
    \centering
    \includegraphics[width=.92\linewidth, trim={0cm .5cm 0cm 0cm}, clip]{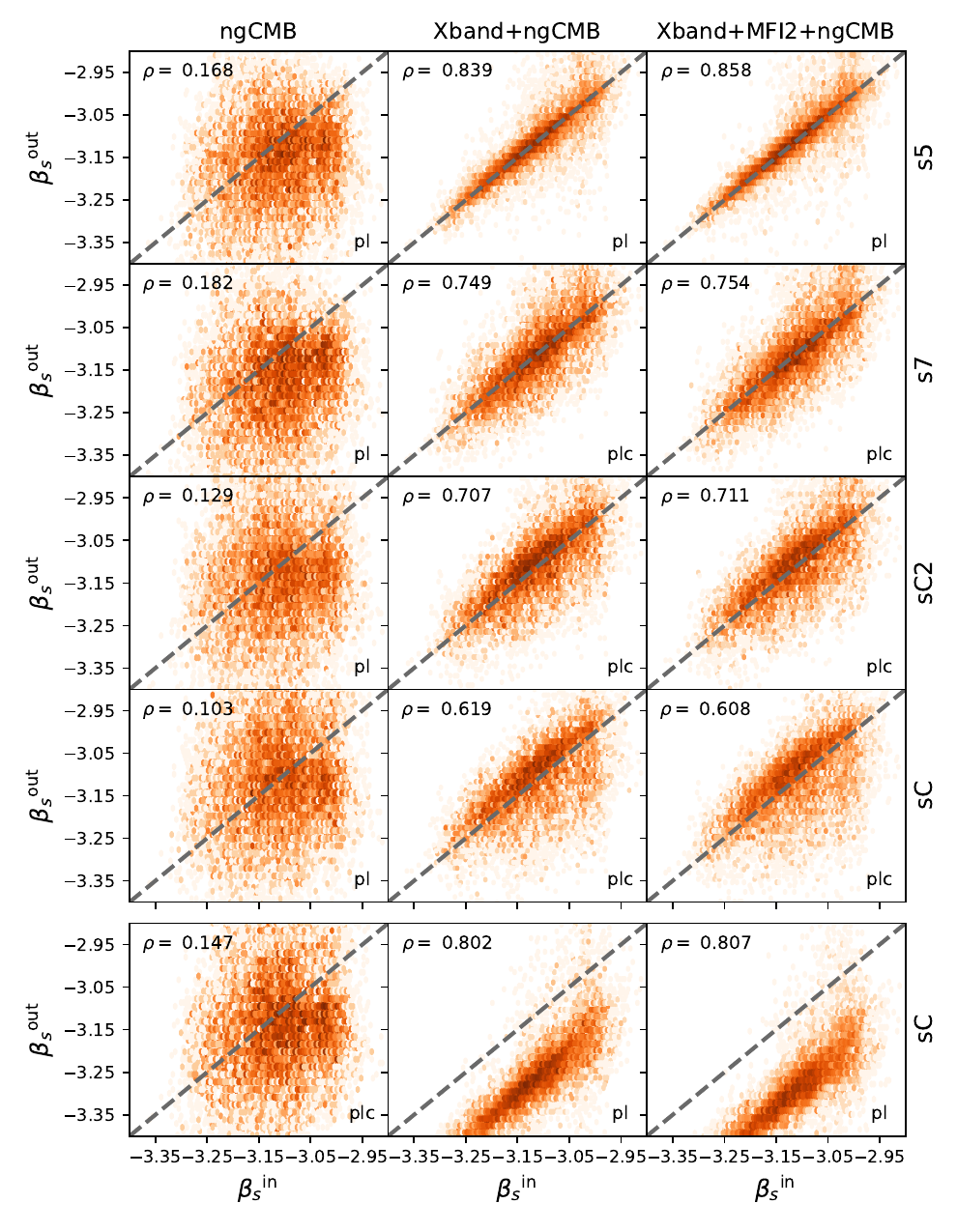}
    \caption{Recovered $\beta_s$ values vs. input $\beta_s$ values for all observable pixels from the Atacama region. The columns, from left to right, correspond to results for ngCMB, Xband+ngCMB, and Xband+MFI2+ngCMB. Rows $1$–$4$ present the outcomes for skies simulated with \texttt{s5}, \texttt{s7}, \texttt{sC2}, and \texttt{sC2} (using the synchrotron model yielding the best $\chi^2_{\rm red}$ map), respectively. Row $5$ shows the results for \texttt{sC} with the alternative (incorrect) synchrotron model. The dashed diagonal line represents $\beta_s^{\rm out} = \beta_s^{\rm in}$. The correlation coefficients ($\rho$) are displayed in the upper-left corner of each plot. }
    \label{fig:tt_plots_true_model}
\end{figure*}

\begin{figure*}
    \centering
    \includegraphics[width=\linewidth, trim={2.5cm 0cm 3.25cm 1.5cm}, clip]{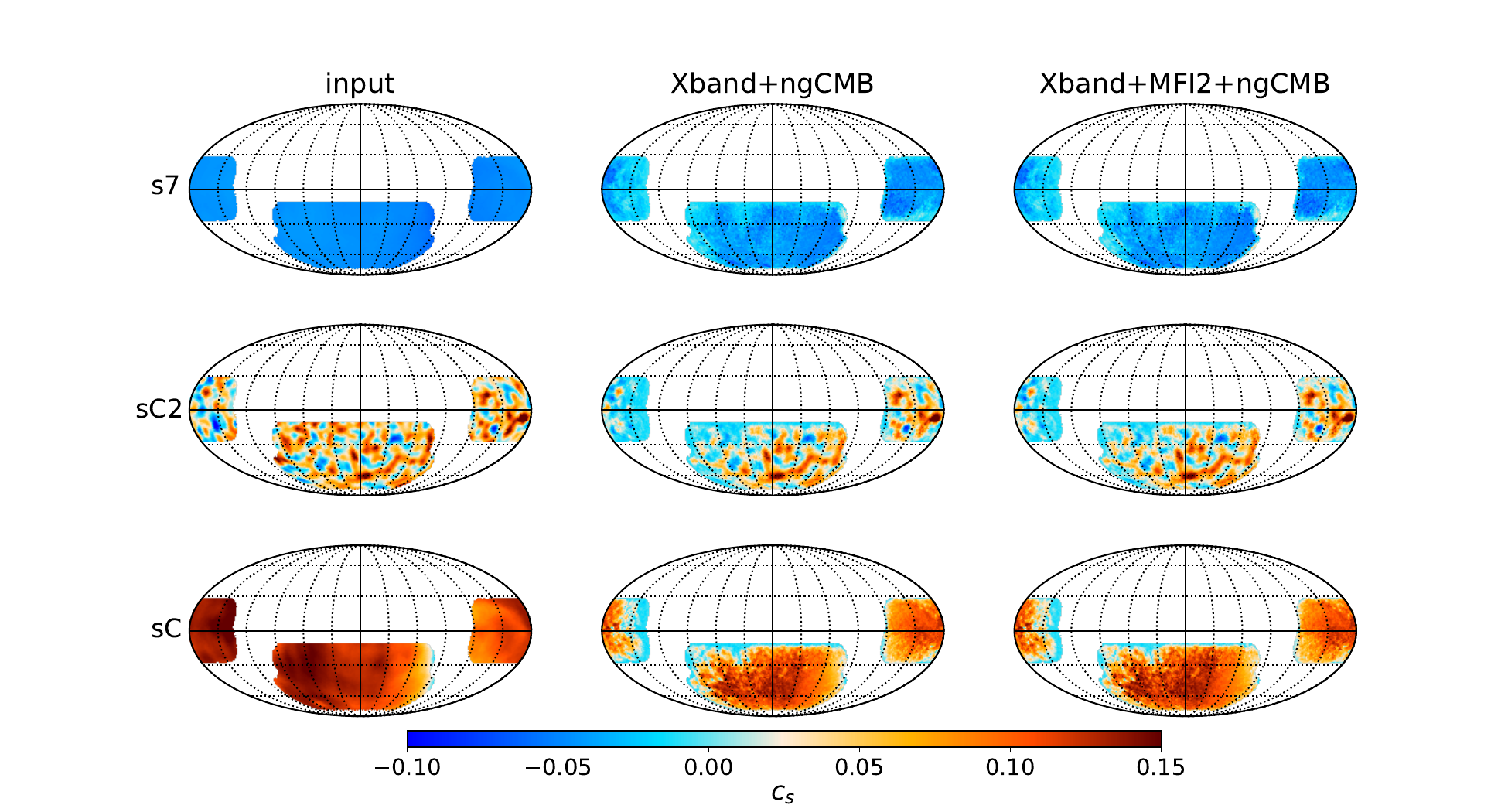}
    \caption{$c_s$ maps: The input template downgraded to $N_{\rm side} = 64$ (left column), recovered $c_s$ for Xband+ngCMB (middle column), and recovered $c_s$ for Xband+MFI2+ngCMB (right column). Results for the \texttt{s7}, \texttt{sC2}, and \texttt{sC} skies are presented in the top, middle, and bottom rows, respectively.}
    \label{fig:c_s_maps}
\end{figure*}

\subsubsection{Spectral Parameters Maps}
\label{subsubsec:spec_par}

We compare the recovered $\beta_s$ maps for the four different synchrotron scenarios considered, using data from ngCMB, Xband+ngCMB, and Xband+MFI2+ngCMB, as shown in Fig.~\ref{fig:beta_s}. The presented results are based on the synchrotron model that yielded the best reduced chi-squared ($\chi^{2}_{\rm red}$) map. This approach closely replicates the procedure applied to real data, where the true underlying synchrotron SED is unknown.

For ngCMB data, we show results obtained by fitting the synchrotron using a power-law model. In the cases of Xband+ngCMB and Xband+MFI2+ngCMB, a power-law model is used for \texttt{s5}, while a power-law model with curvature is applied for \texttt{s7}, \texttt{sC}, and \texttt{sC2}. This distinction reflects the need for more complex models when additional frequency bands are included to account for the increased sensitivity to synchrotron SED variations.

From the ngCMB maps, it is evident that, except in regions with high signal-to-noise ratio (see Fig.~\ref{fig:pol_int_synch_amp_27GHz}), the recovered $\beta_s$ values tend to align closely with the prior set on $\beta_s$, specifically $\beta_s = -3.1$. The dependence of the ngCMB results on prior assumptions is further illustrated in Fig.~\ref{fig:tt_plots_true_model}, which shows a density plot of the recovered $\beta^{\rm out}_s$ values against the downgraded $\beta^{\rm in}s$ template at $N_{\rm side} = 64$. The density plot reveals a horizontal pattern, with the highest density regions clustering around the prior's most probable value ($\beta_s \sim -3.1$) and gradually decreasing in density farther from this value. This indicates that the recovered $\beta_s$ distribution is heavily influenced by the prior $\mathcal{N}(-3.1, 0.1)$, as the data lack sufficient signal-to-noise to independently constrain $\beta_s$.

It is important to note that these results are close to the true $\beta_s$ values primarily because the chosen prior closely approximates the actual distribution. However, this is not guaranteed when working with real data. If the prior's expected value deviates significantly from the true distribution, the bulk density would no longer align with the line $\beta_s^{\rm in} = \beta_s^{\rm out}$. Moreover, due to this strong dependence on the prior, the recovered $\beta_s$ values remain largely similar across all considered skies, with notable differences only in regions of high signal-to-noise.

In high signal-to-noise regions, the recovered $\beta_s$ values better reflect the effective exponent of the power law ($\beta_{\rm eff}(\nu)$) at the lowest ngCMB frequency. Specifically, in the \texttt{s5} scenario, the effective exponent is $\beta_s$, while for the other skies, it is $\beta_s + c_s\log(27\,\mathrm{GHz} / 23\,\mathrm{GHz})$. Consequently, only in the \texttt{s5} case do the recovered $\beta_s$ values closely match the input. For other scenarios, the recovered $\beta_s$ is either steeper or flatter than the input, depending on whether the curvature is negative or positive, see Fig~\ref{fig:c_s_maps}.

This trend changes when we incorporate information from low-frequency receivers. As shown in Fig.~\ref{fig:beta_s} and Fig.~\ref{fig:c_s_maps}, both $\beta_s$ and $c_s$ are more accurately estimated when data from ELFS/SA are included. Deviations from the input values are only observed in the \texttt{s7}, \texttt{sC2}, and \texttt{sC} cases, where a power law with curvature is used to model the synchrotron emission. These deviations occur primarily in regions with very low S/N. This behavior is particularly evident in Fig.~\ref{fig:c_s_maps}, where the estimated $c_s^{\rm out}$ values approach zero, corresponding to the most probable value of the prior distribution. In these low S/N regions, the curvature parameter becomes unconstrained by the data, leading to a recovery of the prior value, $c_s \sim 0$, and a $\beta_s$ value close to the effective exponent $\beta_{\rm eff}(\nu)$ at the lowest analyzed frequency, $\nu = 6.3$,GHz.

Outside of these regions, the additional information provided by the low-frequency receivers allows for the recovery of the true parameters across the rest of the sky. This is evident in Fig.~\ref{fig:tt_plots_true_model}, where $\beta^{\rm out}_s$ shows a strong correlation with $\beta_s^{\rm in}$, with correlation coefficients ranging from $0.6$ to $0.86$ for the various skies analyzed. It is worth noting that the scatter of points increases for the \texttt{s7}, \texttt{sC}, and \texttt{sC2} cases compared to \texttt{s5}, due to the inclusion of the curvature parameter in the fitting model.

\begin{figure*}
    \centering
    \includegraphics[width=\linewidth, trim={.25cm 0.5cm 0cm 0cm}, clip]{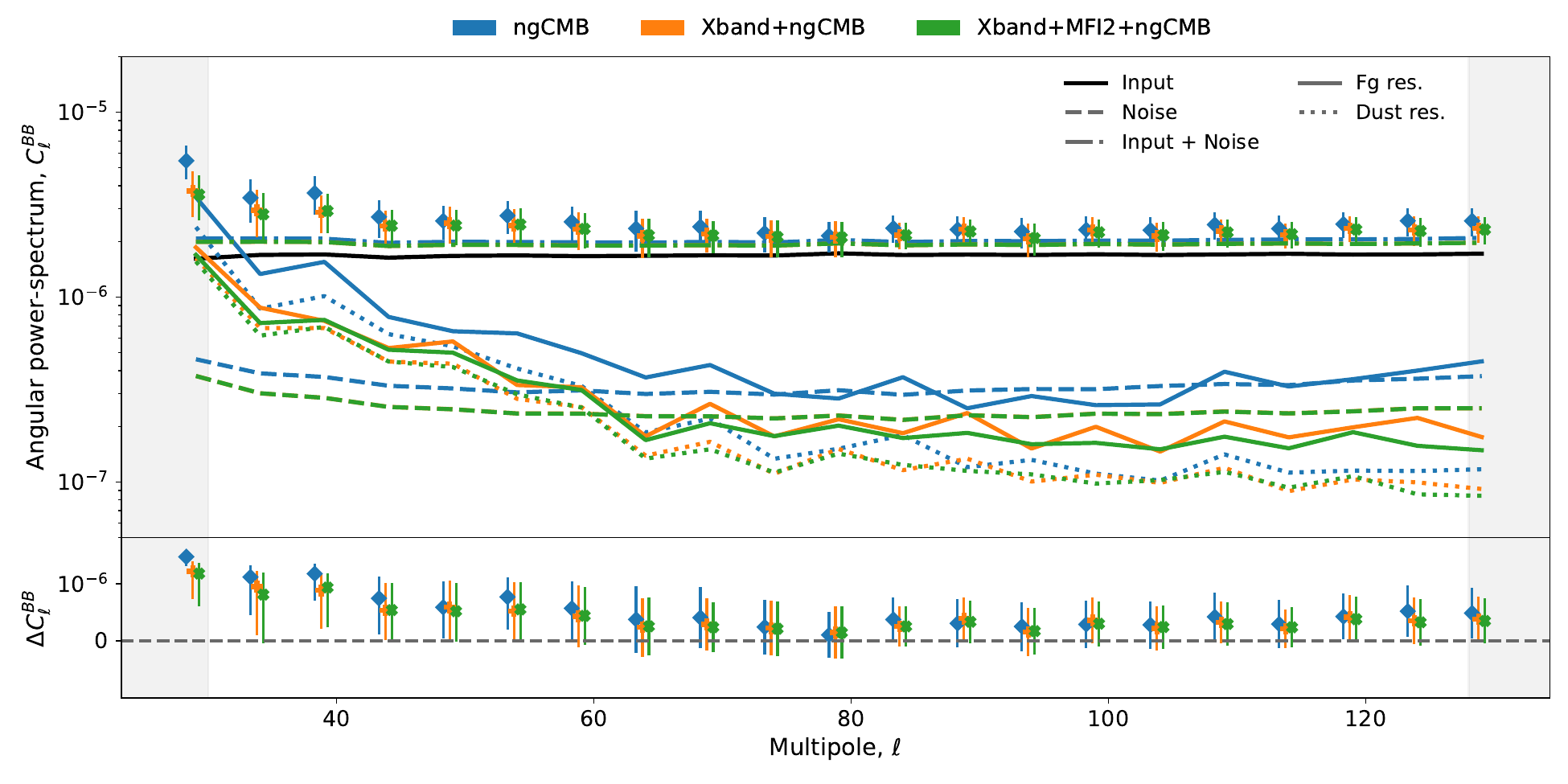}
    \caption{\textit{Top panel}: $BB$ angular power spectra derived using ngCMB (blue), Xband+ngCMB (orange), and Xband+MFI2+ngCMB (green). The recovered CMB is shown as error bars, while the foreground, dust, and noise residuals are represented by solid, dotted, and dashed colored lines, respectively. The theoretical CMB power spectrum is shown as a solid black line. \textit{Bottom panel}: The difference between the recovered CMB angular power spectra and the input CMB power spectra, which corresponds to the theoretical CMB + noise power spectrum. Gray shadowed areas are not included in the analysis to estimate $r$.}
    \label{fig:bb_powspec}
\end{figure*}

For the \texttt{sC} sky, we also present results obtained when fitting the data using the alternative model: a power law with curvature for ngCMB data alone, and a power law for ngCMB combined with ELFS/SA, as shown in the last row of Fig.~\ref{fig:tt_plots_true_model}. As expected, ngCMB data alone cannot constrain the curvature parameter, resulting in recovered values consistent with zero, which aligns with the prior expectation. Consequently, the $\beta_s^{\rm out}$ values are similar to those obtained when fitting the synchrotron using a power law model, leading to comparable distributions of points in the $4^{\rm th}$ and $5^{\rm th}$ rows of Fig.~\ref{fig:tt_plots_true_model}. When low-frequency receivers are incorporated into the analysis, the correlation coefficient $\rho$ is found to be higher when a power law model is used rather than a power law with curvature. However, in this case, $\beta_s^{\rm out}$ no longer aligns with $\beta_s^{\rm in}$ but instead correlates with $\beta_{\rm eff}(6.3,\mathrm{GHz})$.

In summary, ngCMB data alone can only constrain $\beta_{\rm eff}(27,\mathrm{GHz})$ in regions with high S/N. However, the inclusion of ELFS/SA allows us to probe the underlying synchrotron SED and accurately recover the spectral parameters across the sky, except for small regions with minimal synchrotron emission. Recovering the correct spectral parameters in these low-emission regions is not critical for an unbiased estimation of the tensor-to-scalar ratio, as the negligible synchrotron emission in these areas does not significantly affect the cleaning process.

\subsection{Primordial Gravitational Waves}
\label{subsec:cosmology}

In this section, we assess how the additional data from ELFS/SA enhances ngCMB's capability to detect or constrain PGWs. Section~\ref{subsubsec:bb_powspec} compares the foreground and noise residuals when using ngCMB, Xband+ngCMB, and Xband+MFI2+ngCMB data. Section~\ref{subsubsec:r} presents the results for the tensor-to-scalar ratio, highlighting the impact of incorporating Xband and MFI2 data.

\subsubsection{Foreground and Noise Residuals}
\label{subsubsec:bb_powspec}

Here, we evaluate how improved recovery of spectral parameters translates into reduced foreground residuals at the power spectrum level. Figure~\ref{fig:bb_powspec} presents the average \textit{BB} angular power spectra for total foreground residuals (solid colored lines), residuals from thermal dust alone (dotted colored lines), and statistical noise after component separation (dashed coloured lines). The results correspond to a synchrotron sky simulated using the \texttt{s5} model\footnote{Similar results are found for the other skies considered.}, with blue, orange, and green lines representing ngCMB, Xband+ngCMB, and Xband+MFI2+ngCMB, respectively.

As shown in Fig.~\ref{fig:bb_powspec}, for ngCMB alone, total foreground residuals contain more power than the thermal dust residuals at large angular scales, indicating that synchrotron residuals at these scales are not negligible. However, when ELFS/SA data are included, the total foreground residuals decrease, with dust residuals becoming the dominant component due to the improved characterization of synchrotron spectral indices. Moreover, the inclusion of low-frequency information slightly improves the dust residuals, thanks to the correlation between dust and synchrotron components. In other words, adding low-frequency data provides a complementary and cost-effective way to enhance the characterization of high-frequency components.

At intermediate scales ($\ell \gtrsim 70$), synchrotron residuals become the dominant source of error, as they decay more slowly with increasing multipole compared to thermal dust residuals. However, the inclusion of ELFS/SA reduces the total foreground residuals, even though synchrotron residuals are no longer negligible relative to thermal dust residuals. This is partially due to the lower-resolution analysis ($N_{\rm side} = 64$) and the beam convolution to $90$\,arcmin, which introduce some degree of mismodeling, particularly at smaller scales.

Despite this, these smaller scales are less critical for constraining the tensor-to-scalar ratio ($r$), meaning that synchrotron residuals have a limited impact on $r$ estimation. For example, repeating the analysis with a tenfold increase in the number of X-band detectors (resulting in a sensitivity improvement by a factor of $\sqrt{10}$) reduces foreground residuals in the $\ell$-band $60$–$128$ by an average of $26$\%. However, this does not improve the bias in the recovered tensor-to-scalar ratio. While the uncertainty on $r$ slightly decreases, the recovered value remains close to the original because the dominant contamination stems from thermal dust. Furthermore, increasing the number of detectors by a factor of $10$ does not significantly improve dust characterization compared to the baseline Xband case, as the correlation between dust and synchrotron can only help up to a certain point.

\subsubsection{Tensor-to-scalar ratio}
\label{subsubsec:r}

\begin{figure}
    \centering
    \includegraphics[width=.75\linewidth, trim={.5cm .5cm 1cm .5cm}, clip]{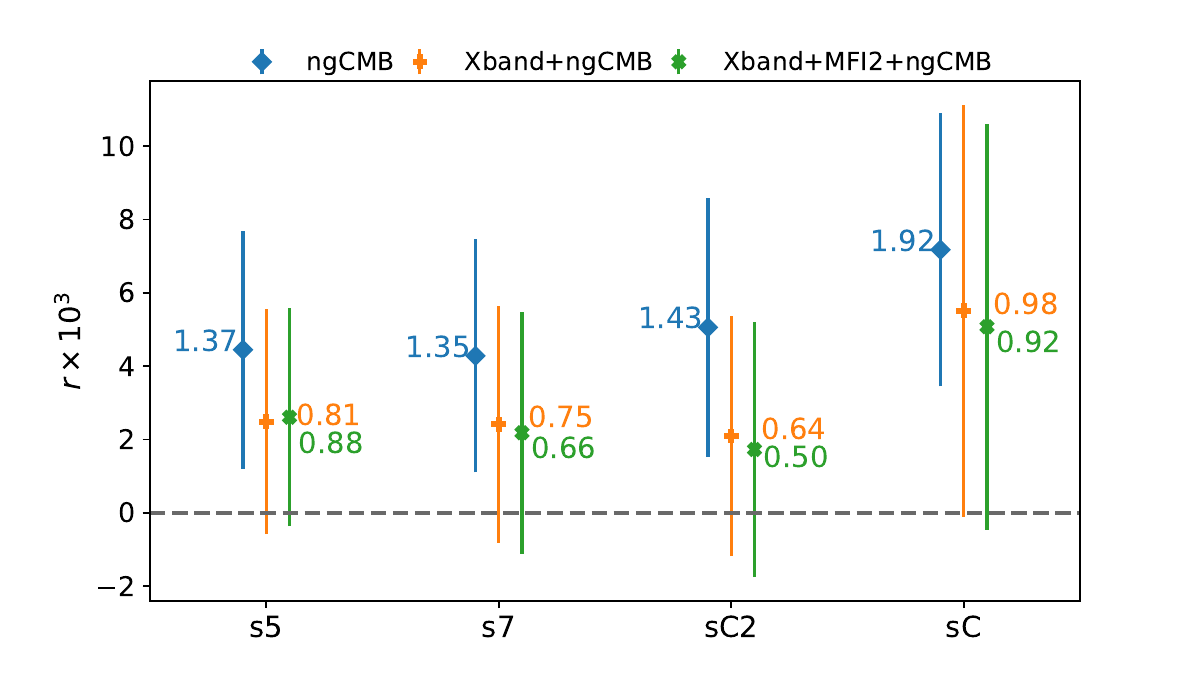}
    \caption{Error bar plot showing the average and standard deviation of the recovered $r$ values for ngCMB (blue), Xband+ngCMB (orange), and Xband+MFI2+ngCMB (green). Results are presented for each synchrotron sky model (\texttt{s5}, \texttt{s7}, \texttt{sC2}, \texttt{sC}). The number displayed next to each point indicates how many standard deviations the average is from the true value. The bias observed in the ngCMB case can be removed using techniques like foreground marginalization as shown in \cite{so_forecast_2023}.}
    \label{fig:r_per_sky}
\end{figure}

In this section, we present the tensor-to-scalar ratio, $r$, obtained using either ngCMB data alone or ngCMB combined with ELFS/SA for the four sky models considered. Fig.~\ref{fig:r_per_sky} displays the average values of $r$ and the associated standard deviations, derived from 50 simulations for each sky and experimental setup. The $r$ values correspond to those obtained from the best-fit models in the $\chi^2_{\rm red}$ maps.

First, we observe that the estimated $r$ values are slightly biased toward positive values due to the non-negligible residuals from thermal dust. However, while the results for Xband+ngCMB and Xband+MFI2+ngCMB are consistent with the true value within one sigma, the ngCMB-alone case often shows a detection with 1–2 sigma significance on average.

The SO collaboration also found that the \texttt{d10s5} model produced biased results when using common but unoptimised component separation techniques \cite{so_forecast_2023}. However, they showed that this bias can be removed by employing techniques such as moment expansion for power-spectrum-based parametric methods or by applying foreground marginalization at the power spectrum level to the results obtained with a map-based parametric method. In this paper, we demonstrate that the bias can also be eliminated by incorporating additional information from low-frequency channels, as illustrated in Fig.~\ref{fig:r_per_sky}. The inclusion of ELFS/SA data therefore potentially removes the need for foreground marginalization which increases $\sigma_r$ and thus reduces the constraining power on PGWs and inflationary models \cite{so_forecast_2023}. The addition of complementary low-frequency data also allows the spectral shape of the foreground residuals to be better constrained which again may mitigate against additional biases in the $r$ estimate if the assumed foregrounds models deviate significantly from the true residuals.

If we compare the performance of Xband+ngCMB versus Xband+MFI2+ngCMB, for the \texttt{s5} sky, we find that the additional data from MFI2 does not reduce the bias compared to Xband+ngCMB, as only two data points are sufficient to characterize a power law. However, when the synchrotron emission deviates from a simple power law, we observe that incorporating data at intermediate frequencies between the Xband and ngCMB enhances the characterization of additional parameters, thereby reducing the bias in the estimation of $r$.

Additionally, we find that for the \texttt{s5} sky, the uncertainty on $r$ decreases by 6\% with the inclusion of Xband and by 8\% when adding Xband+MFI2. Conversely, for \texttt{s7}, \texttt{sC2}, and \texttt{sC} skies, the uncertainty on $r$ increases with the inclusion of Xband and MFI2 due to the introduction of the curvature parameter in the component separation process. It is worth noting that this analysis has not been optimized for minimizing $\sigma_r$ while preserving an unbiased $r$ estimate. Advanced techniques, such as clustering spectral parameters, could reduce $\sigma_r$ by limiting the number of model parameters needed to fit the sky. For instance, assuming that both $\beta_s$ and $c_s$ vary spatially on scales as small as $1^{\circ}$ may be overly conservative. However, care must be taken to avoid incorrect modeling, which could increase bias.

Here, we consider the correct model (aside from potential downgrading effects) to evaluate whether a non-biased $r$ estimate can be recovered in the case of complex synchrotron. We emphasize that our results do not suggest ngCMB cannot impose tighter constraints on $r$ than our estimates; rather, this study aims to evaluate the impact of adding ELFS/SA in the context of complex sky models.

\section{Conclusions} \label{sec:conclusions}

In this study, we have quantified the benefits of incorporating low-frequency data from ELFS/SA for mitigating contamination from diffuse synchrotron emission in CMB observations. Our analysis highlights the following key implications:

\begin{enumerate}
    \item The inclusion of low-frequency data significantly improves the recovery of the synchrotron spectral SED. With the addition of just the Xband data, we can verify whether the assumption of synchrotron emission following a simple power law is valid.
    \item If the synchrotron's SED is more complex than 
    a simple power law both Xband and Xband+MFI2 data enables the accurate recovery of both the synchrotron spectral index $\beta_s$ and curvature parameter $c_s$ across most regions of the sky.
    \item Within the foreground cleaning approach considered in this work, our results show that incorporating ELFS/SA data into ngCMB substantially reduces synchrotron residuals in the recovered maps, particularly at large angular scales where synchrotron contamination is most pronounced
\end{enumerate}

These findings provide support to the opportunity of extending frequency coverage to low frequencies in future CMB experiments. By providing accurate models of synchrotron emission, ELFS data ensures that experiments like \textit{LiteBIRD} and Stage-4 CMB experiments can rely on extra-robustness in the results concerning PGWs coming from more control on the synchrotron contamination. 
Additionally, the improved characterization of synchrotron SEDs can benefit studies of Galactic magnetic fields and the interstellar medium.

\section*{Acknowledgements}

EdlH acknowledges support from IN2P3. R.B. Barreiro, F.J. Casas, E. Martínez-González and P. Vielva would like to acknowledge financial support from the Spanish MCIN/AEI/10.13039/501100011033, project ref. PID2022-139223OB-C21 (funded also by European Union NextGenerationEU/PRTR), and from the Horizon Europe research and innovation program under GA 101135036 (RadioForegroundsPlus). A.C. Taylor and M.E. Jones also acknowledge support from the Horizon Europe Project RadioForegroundsPlus (GA 101135036) which is supported in the UK by UKRI grant number 10101603. 
JARM acknowledges finantial support from the Spanish MCIN/AEI/10.13039/501100011033, project ref. PID2023-151567NB-I00, and from the Horizon Europe research and innovation program under GA 101135036 (RadioForegroundsPlus).
This is not an official SO Collaboration paper. This research used resources from the National Energy Research Scientific Computing Center (NERSC), a U.S. Department of Energy Office of Science User Facility. The author thankfully acknowledges the computer resources, technical expertise and assistance provided by the Advanced Computing \& e-Science team at IFCA. We acknowledge the use of the \texttt{healpy}~\cite{healpy}, \texttt{pysm}~\cite{PYSM_software}, \texttt{NaMaster}~\cite{namaster}, \texttt{emcee}~\cite{foreman2013emcee}, \texttt{numpy}~\cite{numpy}, and \texttt{matplotlib}~\cite{matplotlib} software packages.


\bibliographystyle{JHEP}
\bibliography{references}

\providecommand{\href}[2]{#2}\begingroup\raggedright\begin{thebibliography}{10}

\bibitem{guth1981inflationary}
A.H.~{Guth}, \emph{{Inflationary universe: A possible solution to the horizon
  and flatness problems}},
  \href{https://doi.org/10.1103/PhysRevD.23.347}{\emph{\prd} {\bfseries 23}
  (1981) 347}.

\bibitem{linde1982new}
A.D.~{Linde}, \emph{{A new inflationary universe scenario: A possible solution
  of the horizon, flatness, homogeneity, isotropy and primordial monopole
  problems}}, \href{https://doi.org/10.1016/0370-2693(82)91219-9}{\emph{Physics
  Letters B} {\bfseries 108} (1982) 389}.

\bibitem{fg_tensor_to_scalar_ratio}
J.~{Errard}, S.M.~{Feeney}, H.V.~{Peiris} and A.H.~{Jaffe}, \emph{{Robust
  forecasts on fundamental physics from the foreground-obscured,
  gravitationally-lensed CMB polarization}},
  \href{https://doi.org/10.1088/1475-7516/2016/03/052}{\emph{\jcap} {\bfseries
  2016} (2016) 052} [\href{https://arxiv.org/abs/1509.06770}{{\ttfamily
  1509.06770}}].

\bibitem{bicep_nondetection_r}
{BICEP2 Collaboration}, P.A.R.~{Ade}, R.W.~{Aikin}, D.~{Barkats},
  S.J.~{Benton}, C.A.~{Bischoff} et~al., \emph{{Detection of B-Mode
  Polarization at Degree Angular Scales by BICEP2}},
  \href{https://doi.org/10.1103/PhysRevLett.112.241101}{\emph{\prl} {\bfseries
  112} (2014) 241101} [\href{https://arxiv.org/abs/1403.3985}{{\ttfamily
  1403.3985}}].

\bibitem{bicep_planck_dust_r}
{BICEP2/Keck and Planck Collaborations}, P.A.R.~{Ade}, N.~{Aghanim},
  Z.~{Ahmed}, R.W.~{Aikin}, K.D.~{Alexander} et~al., \emph{{Joint Analysis of
  BICEP2/Keck Array and Planck Data}},
  \href{https://doi.org/10.1103/PhysRevLett.114.101301}{\emph{\prl} {\bfseries
  114} (2015) 101301} [\href{https://arxiv.org/abs/1502.00612}{{\ttfamily
  1502.00612}}].

\bibitem{dust_mismodelling_remazeilles}
M.~{Remazeilles}, C.~{Dickinson}, H.K.K.~{Eriksen} and I.K.~{Wehus},
  \emph{{Sensitivity and foreground modelling for large-scale cosmic microwave
  background B-mode polarization satellite missions}},
  \href{https://doi.org/10.1093/mnras/stw441}{\emph{\mnras} {\bfseries 458}
  (2016) 2032} [\href{https://arxiv.org/abs/1509.04714}{{\ttfamily
  1509.04714}}].

\bibitem{dust_mismodelling_hensley}
B.S.~{Hensley} and P.~{Bull}, \emph{{Mitigating Complex Dust Foregrounds in
  Future Cosmic Microwave Background Polarization Experiments}},
  \href{https://doi.org/10.3847/1538-4357/aaa489}{\emph{\apj} {\bfseries 853}
  (2018) 127} [\href{https://arxiv.org/abs/1709.07897}{{\ttfamily
  1709.07897}}].

\bibitem{SO}
P.~{Ade}, J.~{Aguirre}, Z.~{Ahmed}, S.~{Aiola}, A.~{Ali}, D.~{Alonso} et~al.,
  \emph{{The Simons Observatory: science goals and forecasts}},
  \href{https://doi.org/10.1088/1475-7516/2019/02/056}{\emph{\jcap} {\bfseries
  2019} (2019) 056} [\href{https://arxiv.org/abs/1808.07445}{{\ttfamily
  1808.07445}}].

\bibitem{LiteBIRD_ptep}
{LiteBIRD Collaboration}, E.~{Allys}, K.~{Arnold}, J.~{Aumont}, R.~{Aurlien},
  S.~{Azzoni} et~al., \emph{{Probing cosmic inflation with the LiteBIRD cosmic
  microwave background polarization survey}},
  \href{https://doi.org/10.1093/ptep/ptac150}{\emph{Progress of Theoretical and
  Experimental Physics} {\bfseries 2023} (2023) 042F01}
  [\href{https://arxiv.org/abs/2202.02773}{{\ttfamily 2202.02773}}].

\bibitem{CMB-S4}
K.~{Abazajian}, G.~{Addison}, P.~{Adshead}, Z.~{Ahmed}, S.W.~{Allen},
  D.~{Alonso} et~al., \emph{{CMB-S4 Science Case, Reference Design, and Project
  Plan}}, \href{https://doi.org/10.48550/arXiv.1907.04473}{\emph{arXiv
  e-prints} (2019) arXiv:1907.04473}
  [\href{https://arxiv.org/abs/1907.04473}{{\ttfamily 1907.04473}}].

\bibitem{MFI}
J.A.~{Rubi{\~n}o-Mart{\'\i}n}, F.~{Guidi}, R.T.~{G{\'e}nova-Santos},
  S.E.~{Harper}, D.~{Herranz}, R.J.~{Hoyland} et~al., \emph{{QUIJOTE scientific
  results - IV. A northern sky survey in intensity and polarization at 10-20
  GHz with the multifrequency instrument}},
  \href{https://doi.org/10.1093/mnras/stac3439}{\emph{\mnras} {\bfseries 519}
  (2023) 3383} [\href{https://arxiv.org/abs/2301.05113}{{\ttfamily
  2301.05113}}].

\bibitem{SPASS}
E.~{Carretti}, M.~{Haverkorn}, L.~{Staveley-Smith}, G.~{Bernardi},
  B.M.~{Gaensler}, M.J.~{Kesteven} et~al., \emph{{S-band Polarization All-Sky
  Survey (S-PASS): survey description and maps}},
  \href{https://doi.org/10.1093/mnras/stz806}{\emph{\mnras} {\bfseries 489}
  (2019) 2330} [\href{https://arxiv.org/abs/1903.09420}{{\ttfamily
  1903.09420}}].

\bibitem{C-BASSPol}
A.C.~{Taylor}, \emph{{The C-Band All-Sky Survey}},
  \href{https://doi.org/10.48550/arXiv.1805.05484}{\emph{arXiv e-prints} (2018)
  arXiv:1805.05484} [\href{https://arxiv.org/abs/1805.05484}{{\ttfamily
  1805.05484}}].

\bibitem{planck_int_low_diff_fg}
{Planck Collaboration}, P.A.R.~{Ade}, N.~{Aghanim}, M.I.R.~{Alves},
  M.~{Arnaud}, M.~{Ashdown} et~al., \emph{{Planck 2015 results. XXV. Diffuse
  low-frequency Galactic foregrounds}},
  \href{https://doi.org/10.1051/0004-6361/201526803}{\emph{\aap} {\bfseries
  594} (2016) A25} [\href{https://arxiv.org/abs/1506.06660}{{\ttfamily
  1506.06660}}].

\bibitem{synch_galprop_model}
E.~{Orlando} and A.~{Strong}, \emph{{Galactic synchrotron emission with cosmic
  ray propagation models}},
  \href{https://doi.org/10.1093/mnras/stt1718}{\emph{\mnras} {\bfseries 436}
  (2013) 2127} [\href{https://arxiv.org/abs/1309.2947}{{\ttfamily 1309.2947}}].

\bibitem{synchrotron_aging}
M.J.~{Siah} and P.J.~{Wiita}, \emph{{Synchrotron Aging in Radio Sources. I.
  Spatial Variations in Radio Lobes}},
  \href{https://doi.org/10.1086/169353}{\emph{\apj} {\bfseries 363} (1990)
  411}.

\bibitem{synchrotron_self_absorption}
R.A.~{Chevalier}, \emph{{Synchrotron Self-Absorption in Radio Supernovae}},
  \href{https://doi.org/10.1086/305676}{\emph{\apj} {\bfseries 499} (1998)
  810}.

\bibitem{ELFS}
A.~{Mennella}, K.~{Arnold}, S.~{Azzoni}, C.~{Baccigalupi}, A.~{Banday},
  R.B.~{Barreiro} et~al., \emph{{The European Low Frequency Survey}},
  \href{https://doi.org/10.48550/arXiv.2310.16509}{\emph{arXiv e-prints} (2023)
  arXiv:2310.16509} [\href{https://arxiv.org/abs/2310.16509}{{\ttfamily
  2310.16509}}].

\bibitem{SA}
N.~{Stebor}, P.~{Ade}, Y.~{Akiba}, C.~{Aleman}, K.~{Arnold}, C.~{Baccigalupi}
  et~al., \emph{{The Simons Array CMB polarization experiment}},  in
  \emph{Millimeter, Submillimeter, and Far-Infrared Detectors and
  Instrumentation for Astronomy VIII}, W.S.~{Holland} and J.~{Zmuidzinas},
  eds., vol.~9914 of \emph{Society of Photo-Optical Instrumentation Engineers
  (SPIE) Conference Series}, p.~99141H, July, 2016,
  \href{https://doi.org/10.1117/12.2233103}{DOI}.

\bibitem{model_radio_10MHz_100GHz}
A.~{de Oliveira-Costa}, M.~{Tegmark}, B.M.~{Gaensler}, J.~{Jonas},
  T.L.~{Landecker} and P.~{Reich}, \emph{{A model of diffuse Galactic radio
  emission from 10 MHz to 100 GHz}},
  \href{https://doi.org/10.1111/j.1365-2966.2008.13376.x}{\emph{\mnras}
  {\bfseries 388} (2008) 247}
  [\href{https://arxiv.org/abs/0802.1525}{{\ttfamily 0802.1525}}].

\bibitem{model_radio_45_408MHz}
A.E.~{Guzm{\'a}n}, J.~{May}, H.~{Alvarez} and K.~{Maeda}, \emph{{All-sky
  Galactic radiation at 45 MHz and spectral index between 45 and 408 MHz}},
  \href{https://doi.org/10.1051/0004-6361/200913628}{\emph{\aap} {\bfseries
  525} (2011) A138} [\href{https://arxiv.org/abs/1011.4298}{{\ttfamily
  1011.4298}}].

\bibitem{Orlando2018}
E.~{Orlando}, \emph{{Imprints of cosmic rays in multifrequency observations of
  the interstellar emission}},
  \href{https://doi.org/10.1093/mnras/stx3280}{\emph{\mnras} {\bfseries 475}
  (2018) 2724} [\href{https://arxiv.org/abs/1712.07127}{{\ttfamily
  1712.07127}}].

\bibitem{Orlando2013}
E.~{Orlando} and A.~{Strong}, \emph{{Galactic synchrotron emission with cosmic
  ray propagation models}},
  \href{https://doi.org/10.1093/mnras/stt1718}{\emph{\mnras} {\bfseries 436}
  (2013) 2127} [\href{https://arxiv.org/abs/1309.2947}{{\ttfamily 1309.2947}}].

\bibitem{PlanckCompSep2015}
{Planck Collaboration}, R.~{Adam}, P.A.R.~{Ade}, N.~{Aghanim}, M.I.R.~{Alves},
  M.~{Arnaud} et~al., \emph{{Planck 2015 results. X. Diffuse component
  separation: Foreground maps}},
  \href{https://doi.org/10.1051/0004-6361/201525967}{\emph{\aap} {\bfseries
  594} (2016) A10} [\href{https://arxiv.org/abs/1502.01588}{{\ttfamily
  1502.01588}}].

\bibitem{PlanckLowFreqFg2015}
{Planck Collaboration}, P.A.R.~{Ade}, N.~{Aghanim}, M.I.R.~{Alves},
  M.~{Arnaud}, M.~{Ashdown} et~al., \emph{{Planck 2015 results. XXV. Diffuse
  low-frequency Galactic foregrounds}},
  \href{https://doi.org/10.1051/0004-6361/201526803}{\emph{\aap} {\bfseries
  594} (2016) A25} [\href{https://arxiv.org/abs/1506.06660}{{\ttfamily
  1506.06660}}].

\bibitem{planck_int_diff_fg}
{Planck Collaboration}, R.~{Adam}, P.A.R.~{Ade}, N.~{Aghanim}, M.I.R.~{Alves},
  M.~{Arnaud} et~al., \emph{{Planck 2015 results. X. Diffuse component
  separation: Foreground maps}},
  \href{https://doi.org/10.1051/0004-6361/201525967}{\emph{\aap} {\bfseries
  594} (2016) A10} [\href{https://arxiv.org/abs/1502.01588}{{\ttfamily
  1502.01588}}].

\bibitem{PlanckCompSep2018}
{Planck Collaboration}, Y.~{Akrami}, M.~{Ashdown}, J.~{Aumont},
  C.~{Baccigalupi}, M.~{Ballardini} et~al., \emph{{Planck 2018 results. IV.
  Diffuse component separation}},
  \href{https://doi.org/10.1051/0004-6361/201833881}{\emph{\aap} {\bfseries
  641} (2020) A4} [\href{https://arxiv.org/abs/1807.06208}{{\ttfamily
  1807.06208}}].

\bibitem{pysm3}
{Pan-Experiment Galactic Science Group}, J.~{Borrill}, S.E.~{Clark},
  J.~{Delabrouille}, A.V.~{Frolov}, S.~{Ghosh} et~al., \emph{{Full-sky Models
  of Galactic Microwave Emission and Polarization at Subarcminute Scales for
  the Python Sky Model}},
  \href{https://doi.org/10.3847/1538-4357/adf212}{\emph{\apj} {\bfseries 991}
  (2025) 23} [\href{https://arxiv.org/abs/2502.20452}{{\ttfamily 2502.20452}}].

\bibitem{PSM}
J.~{Delabrouille}, M.~{Betoule}, J.B.~{Melin}, M.A.~{Miville-Desch{\^e}nes},
  J.~{Gonzalez-Nuevo}, M.~{Le Jeune} et~al., \emph{{The pre-launch Planck Sky
  Model: a model of sky emission at submillimetre to centimetre wavelengths}},
  \href{https://doi.org/10.1051/0004-6361/201220019}{\emph{\aap} {\bfseries
  553} (2013) A96} [\href{https://arxiv.org/abs/1207.3675}{{\ttfamily
  1207.3675}}].

\bibitem{beta_s_template}
M.A.~{Miville-Desch{\^e}nes}, N.~{Ysard}, A.~{Lavabre}, N.~{Ponthieu},
  J.F.~{Mac{\'\i}as-P{\'e}rez}, J.~{Aumont} et~al., \emph{{Separation of
  anomalous and synchrotron emissions using WMAP polarization data}},
  \href{https://doi.org/10.1051/0004-6361:200809484}{\emph{\aap} {\bfseries
  490} (2008) 1093} [\href{https://arxiv.org/abs/0802.3345}{{\ttfamily
  0802.3345}}].

\bibitem{WMAP_9years}
C.L.~{Bennett}, D.~{Larson}, J.L.~{Weiland}, N.~{Jarosik}, G.~{Hinshaw},
  N.~{Odegard} et~al., \emph{{Nine-year Wilkinson Microwave Anisotropy Probe
  (WMAP) Observations: Final Maps and Results}},
  \href{https://doi.org/10.1088/0067-0049/208/2/20}{\emph{\apjs} {\bfseries
  208} (2013) 20} [\href{https://arxiv.org/abs/1212.5225}{{\ttfamily
  1212.5225}}].

\bibitem{haslam}
C.G.T.~{Haslam}, U.~{Klein}, C.J.~{Salter}, H.~{Stoffel}, W.E.~{Wilson},
  M.N.~{Cleary} et~al., \emph{{A 408 MHz all-sky continuum survey. I -
  Observations at southern declinations and for the North Polar region.}},
  {\emph{\aap} {\bfseries 100} (1981) 209}.

\bibitem{Curvature_arcade}
A.~{Kogut}, \emph{{Synchrotron Spectral Curvature from 22 MHz to 23 GHz}},
  \href{https://doi.org/10.1088/0004-637X/753/2/110}{\emph{\apj} {\bfseries
  753} (2012) 110} [\href{https://arxiv.org/abs/1205.4041}{{\ttfamily
  1205.4041}}].

\bibitem{synchrotron_spass}
N.~{Krachmalnicoff}, E.~{Carretti}, C.~{Baccigalupi}, G.~{Bernardi},
  S.~{Brown}, B.M.~{Gaensler} et~al., \emph{{S-PASS view of polarized Galactic
  synchrotron at 2.3 GHz as a contaminant to CMB observations}},
  \href{https://doi.org/10.1051/0004-6361/201832768}{\emph{\aap} {\bfseries
  618} (2018) A166} [\href{https://arxiv.org/abs/1802.01145}{{\ttfamily
  1802.01145}}].

\bibitem{synch_mfi}
E.~{\MakeLowercase{D}e la Hoz}, R.B.~{Barreiro}, P.~{Vielva},
  E.~{Mart{\'\i}nez-Gonz{\'a}lez}, J.A.~{Rubi{\~n}o-Mart{\'\i}n},
  B.~{Casaponsa} et~al., \emph{{QUIJOTE scientific results - VIII. Diffuse
  polarized foregrounds from component separation with QUIJOTE-MFI}},
  \href{https://doi.org/10.1093/mnras/stac3020}{\emph{\mnras} {\bfseries 519}
  (2023) 3504} [\href{https://arxiv.org/abs/2301.05117}{{\ttfamily
  2301.05117}}].

\bibitem{mennella-2024a}
A.~{Mennella}, K.~{Arnold}, S.~{Azzoni}, C.~{Baccigalupi}, A.~{Banday},
  R.B.~{Barreiro} et~al., \emph{{The European Low Frequency Survey. Observing
  the radio sky to understand the beginning of the Universe}},  in \emph{mm
  Universe 2023 - Observing the Universe at mm Wavelengths}, vol.~293 of
  \emph{European Physical Journal Web of Conferences}, p.~00031, June, 2024,
  \href{https://doi.org/10.1051/epjconf/202429300031}{DOI}.

\bibitem{MFI2}
R.J.~{Hoyland}, J.A.~{Rubi{\~n}o-Mart{\'\i}n}, M.~{Aguiar-Gonzalez},
  P.~{Alonso-Arias}, E.~{Artal}, M.~{Ashdown} et~al., \emph{{The new
  multi-frequency instrument (MFI2) for the QUIJOTE facility in Tenerife}},  in
  \emph{Millimeter, Submillimeter, and Far-Infrared Detectors and
  Instrumentation for Astronomy XI}, J.~{Zmuidzinas} and J.-R.~{Gao}, eds.,
  vol.~12190 of \emph{Society of Photo-Optical Instrumentation Engineers (SPIE)
  Conference Series}, p.~1219033, Aug., 2022,
  \href{https://doi.org/10.1117/12.2640826}{DOI}.

\bibitem{C-BASS}
M.E.~{Jones}, A.C.~{Taylor}, M.~{Aich}, C.J.~{Copley}, H.C.~{Chiang},
  R.J.~{Davis} et~al., \emph{{The C-Band All-Sky Survey (C-BASS): design and
  capabilities}}, \href{https://doi.org/10.1093/mnras/sty1956}{\emph{\mnras}
  {\bfseries 480} (2018) 3224}
  [\href{https://arxiv.org/abs/1805.04490}{{\ttfamily 1805.04490}}].

\bibitem{2014MNRAS.438.2426K}
O.G.~{King}, M.E.~{Jones}, E.J.~{Blackhurst}, C.~{Copley}, R.J.~{Davis} et~al.,
  \emph{{The C-Band All-Sky Survey (C-BASS): design and implementation of the
  northern receiver}},
  \href{https://doi.org/10.1093/mnras/stt2359}{\emph{MNRAS} {\bfseries 438}
  (2014) 2426} [\href{https://arxiv.org/abs/1310.7129}{{\ttfamily 1310.7129}}].

\bibitem{granet2005}
C.~Granet and G.~James, \emph{Design of corrugated horns: a primer},
  \href{https://doi.org/10.1109/MAP.2005.1487785}{\emph{IEEE Antennas and
  Propagation Magazine} {\bfseries 47} (2005) 76}.

\bibitem{so_forecast_2023}
K.~{Wolz}, S.~{Azzoni}, C.~{Herv{\'\i}as-Caimapo}, J.~{Errard},
  N.~{Krachmalnicoff}, D.~{Alonso} et~al., \emph{{The Simons Observatory:
  Pipeline comparison and validation for large-scale B-modes}},
  \href{https://doi.org/10.1051/0004-6361/202346105}{\emph{\aap} {\bfseries
  686} (2024) A16} [\href{https://arxiv.org/abs/2302.04276}{{\ttfamily
  2302.04276}}].

\bibitem{HEALPix}
K.M.~{G{\'o}rski}, E.~{Hivon}, A.J.~{Banday}, B.D.~{Wandelt}, F.K.~{Hansen},
  M.~{Reinecke} et~al., \emph{{HEALPix: A Framework for High-Resolution
  Discretization and Fast Analysis of Data Distributed on the Sphere}},
  \href{https://doi.org/10.1086/427976}{\emph{\apj} {\bfseries 622} (2005) 759}
  [\href{https://arxiv.org/abs/astro-ph/0409513}{{\ttfamily
  astro-ph/0409513}}].

\bibitem{camb}
A.~{Lewis} and A.~{Challinor}, ``{CAMB: Code for Anisotropies in the Microwave
  Background}.'' Astrophysics Source Code Library, record ascl:1102.026, Feb.,
  2011.

\bibitem{planck_cosmo_par}
{Planck Collaboration}, N.~{Aghanim}, Y.~{Akrami}, M.~{Ashdown}, J.~{Aumont},
  C.~{Baccigalupi} et~al., \emph{{Planck 2018 results. VI. Cosmological
  parameters}}, \href{https://doi.org/10.1051/0004-6361/201833910}{\emph{\aap}
  {\bfseries 641} (2020) A6}
  [\href{https://arxiv.org/abs/1807.06209}{{\ttfamily 1807.06209}}].

\bibitem{dust_mcbride}
L.~{McBride}, P.~{Bull} and B.S.~{Hensley}, \emph{{Characterizing line-of-sight
  variability of polarized dust emission with future CMB experiments}},
  \href{https://doi.org/10.1093/mnras/stac3754}{\emph{\mnras} {\bfseries 519}
  (2023) 4370} [\href{https://arxiv.org/abs/2207.14213}{{\ttfamily
  2207.14213}}].

\bibitem{dust_ritacco}
A.~{Ritacco}, F.~{Boulanger}, V.~{Guillet}, J.-M.~{Delouis}, J.-L.~{Puget},
  J.~{Aumont} et~al., \emph{{Dust polarization spectral dependence from Planck
  HFI data. Turning point for cosmic microwave background
  polarization-foreground modeling}},
  \href{https://doi.org/10.1051/0004-6361/202244269}{\emph{\aap} {\bfseries
  670} (2023) A163} [\href{https://arxiv.org/abs/2206.07671}{{\ttfamily
  2206.07671}}].

\bibitem{ame_review}
C.~{Dickinson}, Y.~{Ali-Ha{\"\i}moud}, A.~{Barr}, E.S.~{Battistelli},
  A.~{Bell}, L.~{Bernstein} et~al., \emph{{The State-of-Play of Anomalous
  Microwave Emission (AME) research}},
  \href{https://doi.org/10.1016/j.newar.2018.02.001}{\emph{\nar} {\bfseries 80}
  (2018) 1} [\href{https://arxiv.org/abs/1802.08073}{{\ttfamily 1802.08073}}].

\bibitem{Haimoud2013}
Y.~{Ali-Ha{\"\i}moud}, \emph{{Spinning Dust Radiation: A Review of the
  Theory}}, \href{https://doi.org/10.1155/2013/462697}{\emph{Advances in
  Astronomy} {\bfseries 2013} (2013) 462697}
  [\href{https://arxiv.org/abs/1211.2748}{{\ttfamily 1211.2748}}].

\bibitem{Draine1999}
B.T.~{Draine} and A.~{Lazarian}, \emph{{Magnetic Dipole Microwave Emission from
  Dust Grains}}, \href{https://doi.org/10.1086/306809}{\emph{\apj} {\bfseries
  512} (1999) 740} [\href{https://arxiv.org/abs/astro-ph/9807009}{{\ttfamily
  astro-ph/9807009}}].

\bibitem{Greaves2018}
J.S.~{Greaves}, A.M.M.~{Scaife}, D.T.~{Frayer}, D.A.~{Green}, B.S.~{Mason} and
  A.M.S.~{Smith}, \emph{{Anomalous microwave emission from spinning
  nanodiamonds around stars}},
  \href{https://doi.org/10.1038/s41550-018-0495-z}{\emph{Nature Astronomy}
  {\bfseries 2} (2018) 662} [\href{https://arxiv.org/abs/1806.04551}{{\ttfamily
  1806.04551}}].

\bibitem{Genova2017}
R.~{G{\'e}nova-Santos}, J.A.~{Rubi{\~n}o-Mart{\'\i}n}, A.~{Pel{\'a}ez-Santos},
  F.~{Poidevin}, R.~{Rebolo}, R.~{Vignaga} et~al., \emph{{QUIJOTE scientific
  results - II. Polarisation measurements of the microwave emission in the
  Galactic molecular complexes W43 and W47 and supernova remnant W44}},
  \href{https://doi.org/10.1093/mnras/stw2503}{\emph{\mnras} {\bfseries 464}
  (2017) 4107} [\href{https://arxiv.org/abs/1605.04741}{{\ttfamily
  1605.04741}}].

\bibitem{Faraday_Rotation}
S.~{Hutschenreuter}, C.S.~{Anderson}, S.~{Betti}, G.C.~{Bower}, J.A.~{Brown},
  M.~{Br{\"u}ggen} et~al., \emph{{The Galactic Faraday rotation sky 2020}},
  \href{https://doi.org/10.1051/0004-6361/202140486}{\emph{\aap} {\bfseries
  657} (2022) A43} [\href{https://arxiv.org/abs/2102.01709}{{\ttfamily
  2102.01709}}].

\bibitem{de2020detection}
E.~{\MakeLowercase{D}e la Hoz}, P.~{Vielva}, R.B.~{Barreiro} and
  E.~{Mart{\'\i}nez-Gonz{\'a}lez}, \emph{{On the detection of CMB B-modes from
  ground at low frequency}},
  \href{https://doi.org/10.1088/1475-7516/2020/06/006}{\emph{\jcap} {\bfseries
  2020} (2020) 006} [\href{https://arxiv.org/abs/2002.12206}{{\ttfamily
  2002.12206}}].

\bibitem{foreman2013emcee}
D.~{Foreman-Mackey}, D.W.~{Hogg}, D.~{Lang} and J.~{Goodman}, \emph{{emcee: The
  MCMC Hammer}}, \href{https://doi.org/10.1086/670067}{\emph{\pasp} {\bfseries
  125} (2013) 306} [\href{https://arxiv.org/abs/1202.3665}{{\ttfamily
  1202.3665}}].

\bibitem{namaster}
D.~{Alonso}, J.~{Sanchez}, A.~{Slosar} and {LSST Dark Energy Science
  Collaboration}, \emph{{A unified pseudo-C$_{{\ensuremath{\ell}}}$
  framework}}, \href{https://doi.org/10.1093/mnras/stz093}{\emph{\mnras}
  {\bfseries 484} (2019) 4127}
  [\href{https://arxiv.org/abs/1809.09603}{{\ttfamily 1809.09603}}].

\bibitem{healpy}
A.~{Zonca}, L.~{Singer}, D.~{Lenz}, M.~{Reinecke}, C.~{Rosset}, E.~{Hivon}
  et~al., \emph{{healpy: equal area pixelization and spherical harmonics
  transforms for data on the sphere in Python}},
  \href{https://doi.org/10.21105/joss.01298}{\emph{The Journal of Open Source
  Software} {\bfseries 4} (2019) 1298}.

\bibitem{PYSM_software}
A.~{Zonca}, B.~{Thorne}, N.~{Krachmalnicoff} and J.~{Borrill}, \emph{{The
  Python Sky Model 3 software}},
  \href{https://doi.org/10.21105/joss.03783}{\emph{The Journal of Open Source
  Software} {\bfseries 6} (2021) 3783}
  [\href{https://arxiv.org/abs/2108.01444}{{\ttfamily 2108.01444}}].

\bibitem{numpy}
C.R.~{Harris}, K.J.~{Millman}, S.J.~{van der Walt}, R.~{Gommers},
  P.~{Virtanen}, D.~{Cournapeau} et~al., \emph{{Array programming with NumPy}},
  \href{https://doi.org/10.1038/s41586-020-2649-2}{\emph{\nat} {\bfseries 585}
  (2020) 357} [\href{https://arxiv.org/abs/2006.10256}{{\ttfamily
  2006.10256}}].

\bibitem{matplotlib}
J.D.~{Hunter}, \emph{{Matplotlib: A 2D Graphics Environment}},
  \href{https://doi.org/10.1109/MCSE.2007.55}{\emph{Comput. Sci. Eng.}
  {\bfseries 9} (2007) 90}.

\end{thebibliography}\endgroup



\appendix
\section{Faraday Rotation} \label{sec:appendix_fr}

\begin{figure*}
        \centering
        \begin{subfigure}{.48\linewidth}
            \includegraphics[width=\linewidth, trim={1.5cm 1.25cm 1.5cm 1.5cm}, clip]{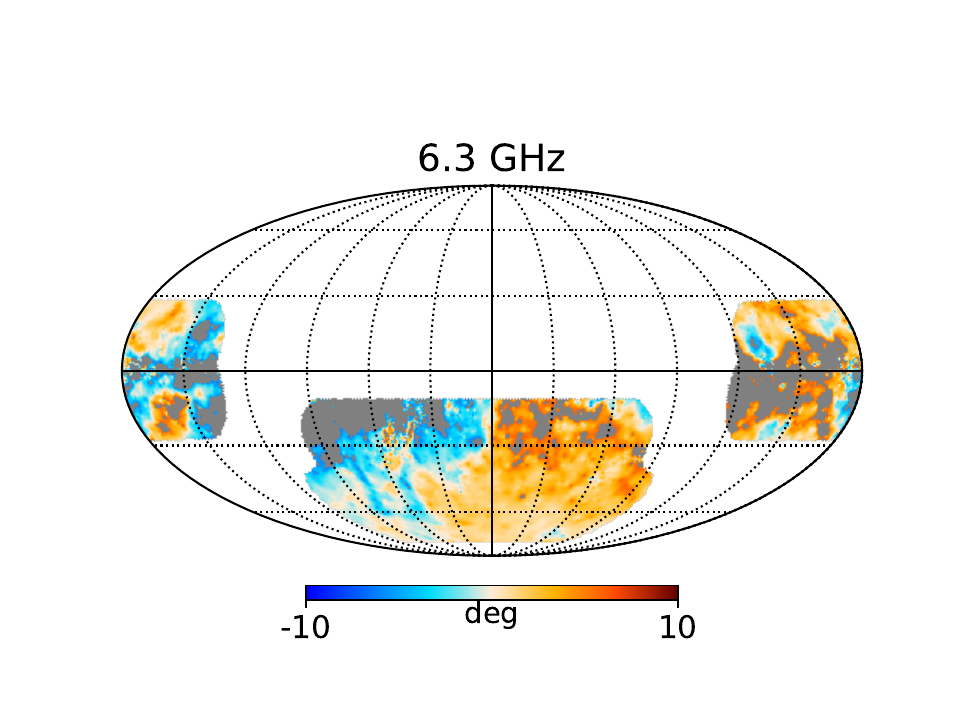}
        \end{subfigure}
        \begin{subfigure}{.48\linewidth}
            \centering
            \includegraphics[width=\linewidth, trim={1.5cm 1.25cm 1.5cm 1.5cm}, clip]{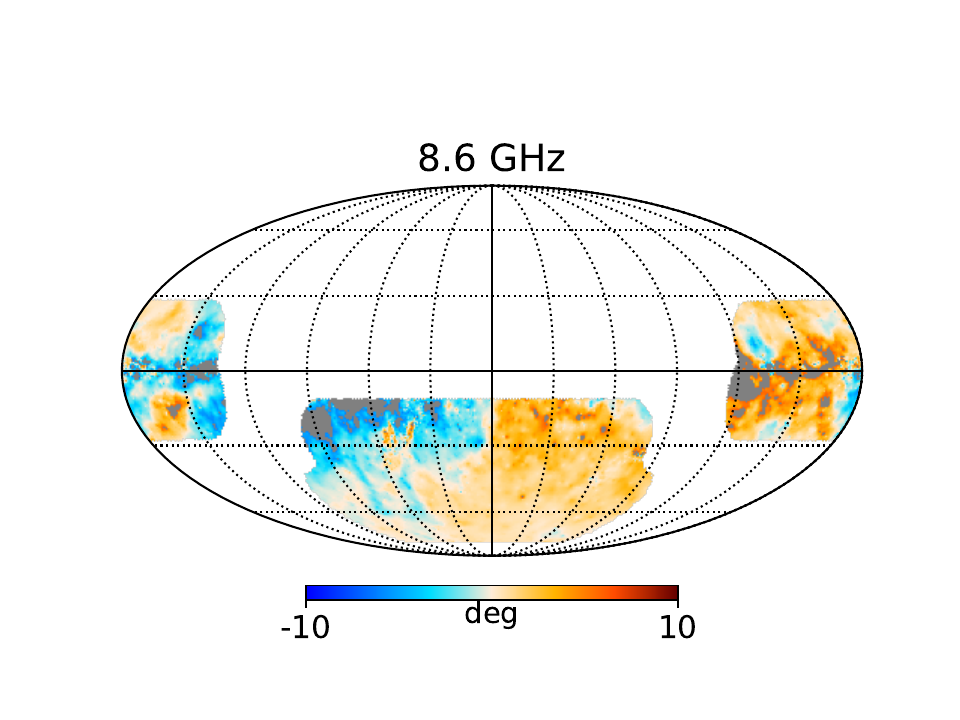}
        \end{subfigure}
        \begin{subfigure}{.48\linewidth}
            \centering
            \includegraphics[width=\linewidth, trim={1.5cm 1.25cm 1.5cm 1.5cm}, clip]{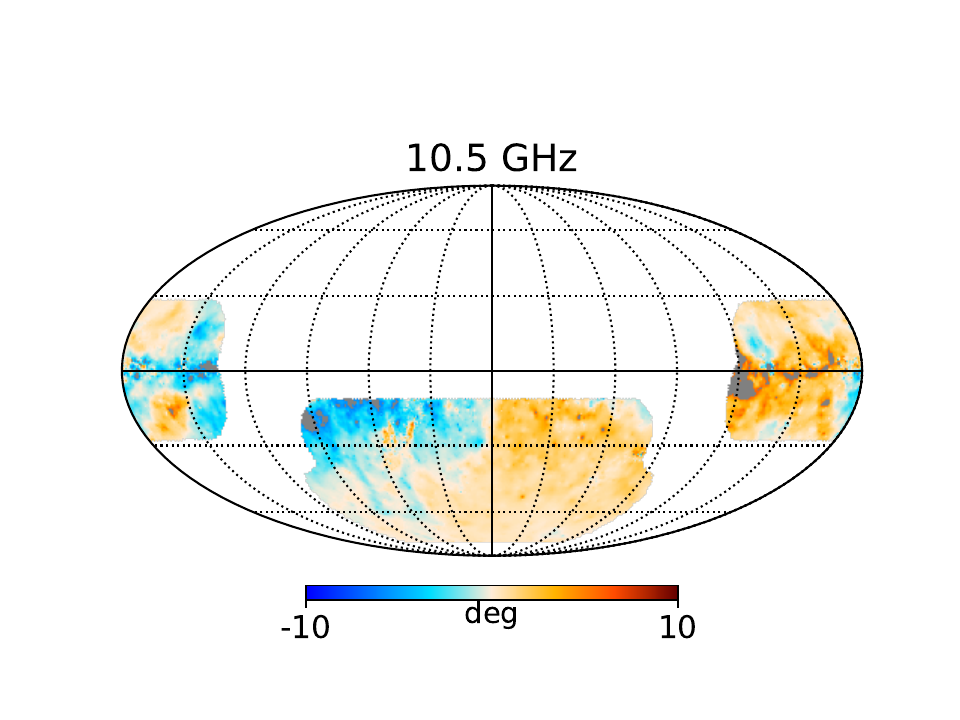}
        \end{subfigure}
        \begin{subfigure}{.48\linewidth}
            \centering
            \includegraphics[width=\linewidth, trim={1.5cm 1.25cm 1.5cm 1.5cm}, clip]{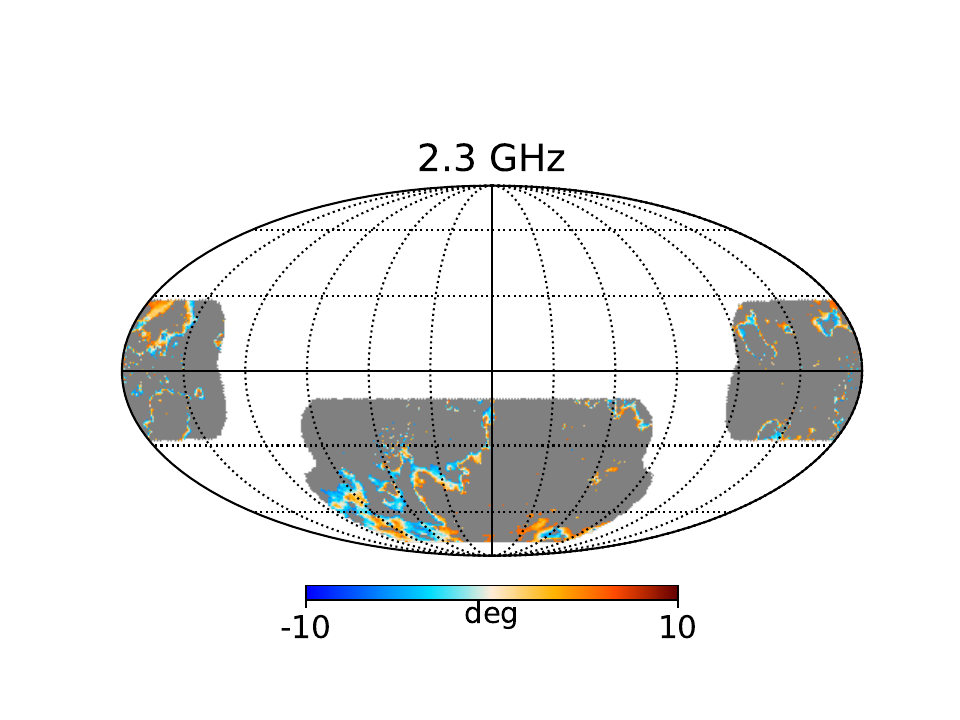}
        \end{subfigure}
    \caption{Estimated Faraday rotation angles for different frequency channels. The model from \cite{Faraday_Rotation} is used to calculate the rotation angles at 6.3 GHz (red), 8.6 GHz (green), and 10.5 GHz (blue) from Xband, as well as at the S-PASS frequency (gray). The gray shaded areas indicate rotation angles greater than $10^\circ$.}
    \label{fig:faraday_rotation}
\end{figure*}

Faraday rotation, the phenomenon where the plane of polarization of synchrotron radiation rotates as it passes through a magnetized medium, can be modeled. However, its impact is often complicated by other effects that mix the $Q$ and $U$ Stokes parameters, making the study of this effect more challenging. This is particularly relevant when combining S-PASS data with data from other ground-based experiments operating from the Southern Hemisphere, as the rotation can distort the measurements.

To address this, the inclusion of the ELFS receiver on the SA telescope, which operates at intermediate frequencies between S-PASS and future ground-based experiments like SO, can help model the Faraday rotation more effectively. This effect scales with the square of the wavelength, meaning its influence is diminished at higher frequencies such as those observed by ELFS/SA.

An estimation of the rotation angles, using the model from \cite{Faraday_Rotation}, is shown in Fig.~\ref{fig:faraday_rotation}. The plot includes the $6.3$, $8.6$, and $10.5$\,GHz channels from Xband, as well as the S-PASS frequency for comparison. The gray shaded areas indicate rotation angles greater than $10^\circ$. For S-PASS, most pixels have values larger than $10^\circ$, while only a few pixels for the $8.6$ and $10.5$\,GHz channels exceed this threshold. The $6.3$\,GHz channel shows more pixels with large rotation angles, but these are primarily in regions with low S/N synchrotron emission.

Thus, the reduced impact of Faraday rotation at higher frequencies than SPASS makes the data easier to combine and potentially be used for a more reliable for joint analysis of SPASS and future ground-based experiments, leading to more accurate results.


\end{document}